\documentclass[aps,eqsecnum]{revtex4}

\usepackage{graphicx}
\usepackage{amsmath}
\usepackage{amsfonts}
\usepackage{mathrsfs}
\usepackage{color}
 \usepackage{slashed}

\usepackage{tikz}
\usepackage{cancel}

\usepackage{amssymb}
\usepackage{pifont}
\newcommand{\xmark}{\ding{53}}%

\def\checkmark{\tikz\fill[scale=0.4](0,.35) -- (.25,0) -- (1,.7) -- (.25,.15) -- cycle;} 

\def\<{\langle}
\def\>{\rangle}

\def\beq{\begin{equation}}
\def\eeq{\end{equation}}
\def\barray{\begin{eqnarray}}
\def\earray{\end{eqnarray}}

\font\numbers=cmss12
\font\upright=cmu10 scaled\magstep1
\def\stroke{\vrule height8pt width0.4pt depth-0.1pt}
\def\topfleck{\vrule height8pt width0.5pt depth-5.9pt}
\def\botfleck{\vrule height2pt width0.5pt depth0.1pt}
\def\Zmath{\vcenter{\hbox{\numbers\rlap{\rlap{Z}\kern
0.8pt\topfleck}\kern 2.2pt
                   \rlap Z\kern 6pt\botfleck\kern 1pt}}}
\def\Qmath{\vcenter{\hbox{\upright\rlap{\rlap{Q}\kern
                   3.8pt\stroke}\phantom{Q}}}}
\def\Nmath{\vcenter{\hbox{\upright\rlap{I}\kern 1.7pt N}}}
\def\Cmath{\vcenter{\hbox{\upright\rlap{\rlap{C}\kern
                   3.8pt\stroke}\phantom{C}}}}
\def\Rmath{\vcenter{\hbox{\upright\rlap{I}\kern 1.7pt R}}}
\def\Z{\ifmmode\Zmath\else$\Zmath$\fi}
\def\Q{\ifmmode\Qmath\else$\Qmath$\fi}
\def\N{\ifmmode\Nmath\else$\Nmath$\fi}
\def\C{\ifmmode\Cmath\else$\Cmath$\fi}
\def\R{\ifmmode\Rmath\else$\Rmath$\fi}

\begin{document}

\title{The Riemann zeros as spectrum and the Riemann hypothesis}

\author{Germ\'an Sierra \vspace{0.2cm} \\ Instituto de F\'isica Te\'orica UAM/CSIC, \\ Universidad Aut\'onoma de Madrid, Cantoblanco, Madrid, Spain.}

\begin{abstract}
We present a spectral realization of the Riemann zeros based on the propagation of a massless Dirac fermion in a region of Rindler spacetime 
and under the action of delta function potentials localized on the square free integers. The corresponding Hamiltonian admits a self-adjoint extension 
that is tuned to the phase of the zeta function, on the critical line, in order to obtain the Riemann zeros as bound states. 
The model suggests a proof of the Riemann hypothesis in the limit where the potentials vanish. 
Finally, we propose an interferometer that may yield an experimental observation of the Riemann zeros.
\end{abstract}

\maketitle

\tableofcontents

\vspace{1cm} 

\section{Introduction} 

One of the most promising approaches to prove the Riemann Hypothesis  \cite{R59}-\cite{C03}
is based on the conjecture, due to P\'olya and Hilbert,  that the Riemann zeros are the eigenvalues of
a quantum mechanical Hamiltonian  \cite{P14}. This bold idea is supported by several  
results and  analogies  involving 
Number Theory, Random Matrix Theory and Quantum Chaos \cite{M74}-\cite{H76}. 
However the construction of a  Hamiltonian whose  spectrum
contains  the Riemann zeros,  has eluded the researchers for several decades. In this paper
we shall review the progress made along this direction starting from the famous $xp$ model proposed in 1999  by Berry,  Keating 
and Connes \cite{BK99}-\cite{C99},  that inspired many works   \cite{A99}-\cite{BB17},
some of them   will be discuss below.  
See \cite{SH11} for a general review  on physical  approaches to the RH.
 Another  approaches to the RH and related material 
 can be found in \cite{PF75}-\cite{ML18b}.

To relate  $xp$ with the Riemann zeros, Berry, Keating and Connes  used  two 
different regularizations. The Berry and Keating regularization
led to a discrete spectrum related to the smooth Riemann zeros  \cite{BK99,BK99b}, while 
Connes's regularization led  to an absorption spectrum where the {\em zeros} are
missing spectral  lines \cite{C99}. 
A physical realization of the Connes model 
was  obtained in 2008 in terms of  the dynamics of an electron moving in two dimensions  under the action of a uniform
perpendicular magnetic field and a electrostatic potential \cite{ST08}. However
this model has not been  able to reproduce the exact  location of the Riemann zeros. 
On the other hand, the Berry-Keating $xp$ model, was revisited in 2011
in terms of the classical Hamiltonians $H = x( p + 1/p)$,  and  $H = (x+1/x)( p + 1/p)$ whose quantizations contain
the smooth approximation of the Riemann zeros \cite{SL11,BK11}. Later on,  these models were generalized
in terms of the family of Hamiltonians $H = U(x) p + V(x)/p$, that were shown to describe the dynamics
of a massive particle in a relativistic spacetime whose metric can be constructed using  the functions $U$ and $V$ 
\cite{S12}. This result  suggested  a  reformulation of 
$H = U(x) p + V(x)/p$ in terms of the massive Dirac equation in the aforementioned spacetimes \cite{MS12}.
Using this reformulation, the Hamiltonian  $H = x(p + 1/p)$ was shown to be equivalent 
to the massive Dirac equation in Rindler spacetime, that 
is the natural arena to study accelerated  observers and the Unruh effect  \cite{S14}. 
This result  provides an appealing   spacetime interpretation of the $xp$ model
and in particular of the smooth Riemann zeros. 

To obtain the exact
{\em zeros},   one has to make   further  modifications  of  the Dirac model.
First of all, the fermion has to become  massless. This change is suggested by a field theory
interpretation of the   P\'olya's  $\xi$ function 
and its comparison with the Riemann's  $\xi$ function. On the other hand,
inspired by the   Berry's conjecture on the relation between prime numbers  and periodic orbits \cite{B86,K99} 
 we  incorporated  the prime numbers 
into the Dirac action by means of  Dirac delta functions \cite{S14}. 
These delta functions represent moving mirrors 
that reflect or  transmit massless fermions. The spectrum of the complete  model 
can be analyzed using transfer matrix techniques that can be solved
exactly in the limit where the reflection amplitudes of the mirrors go to zero, that is when the mirrors
become transparent.  In this limit we find that the  {\em zeros} on the critical line  are eigenvalues
of the Hamiltonian by choosing appropriately the parameter that characterizes the self-adjoint
extension of the Hamiltonian. One obtains in this manner a spectral realization of the Riemann zeros that 
differs from the P\'olya and Hilbert conjecture in the sense that one needs to fine tune a parameter
to {\em see} each individual {\em zero}. In our approach we are not able to find a single Hamiltonian 
encompassing all the {\em zeros} at once. Finally, we  propose an experimental realization of the
Riemann zeros using an interferometer consisting in 
an array of semitransparent mirrors, or beam splitters,   placed at positions related
to the logarithms of the prime numbers.

The paper is organized in a historical  and pedagogical way presenting at the end of each section  a  summary of 
achievements  (\checkmark), shortcomings/obstacles   (\xmark) and  questions/suggestions (?).

\section{The semiclassical $XP$  Berry, Keating and Connes  model} 

\begin{figure}[h!]
\begin{center}
\includegraphics[width=.3\linewidth]{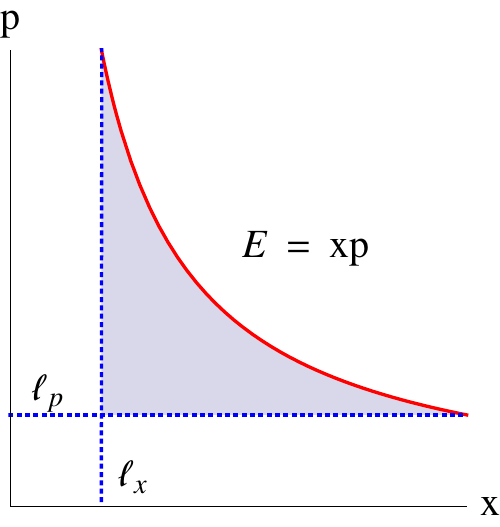}
\hspace{1.5cm} 
\includegraphics[width=.27\linewidth]{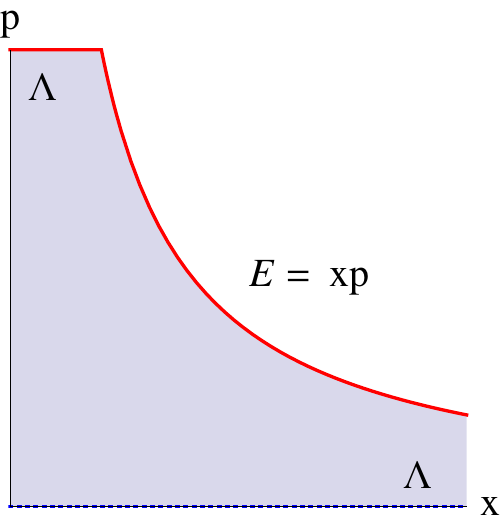}
\end{center}
\caption{Left: The region in shadow  describes the allowed 
phase space with area  $A_{\rm BK}$ 
bounded by the classical  trajectory  (\ref{cl0}) with $E>0$ 
and the constraints  $x \geq  \ell_x, p \geq \ell_p$.  Right: Same as before
with the constraints $0 < x, p  < \Lambda$. 
  }
\label{xp-curve}
\end{figure}

In this section we review the main results concerning the classical and   semiclassical  $xp$ model \cite{BK99,BK99b,C99}. 
A classical trajectory of the Hamiltonian   $H_{} = xp$,
with energy $E$, is   given by 

\beq
x(t) = x_0 \,  e^t, \qquad p(t) = p_0  \, e^{-t},  \qquad  E = x_0 p_0  \, , 
\label{cl0}
\eeq
that  traces  the   parabola $E= x p$  in phase space plotted in Fig.\ref{xp-curve}. 
$E$ has the dimension  of an action, so one should multiply
$xp$ by a frequency to get an  energy, but for the time
being we keep the notation $H=xp$. 
Under a time reversal transformation, $x \rightarrow x, p \rightarrow - p$ one finds 
 $x p \rightarrow - x p$, so  that 
this symmetry is broken. This is  why reversing  the time variable $t$
in  (\ref{cl0}) does  not yield  a trajectory generated by $xp$. 
As $t \rightarrow \infty$, the trajectory becomes unbounded, that is 
$|x| \rightarrow \infty$,  so  one  expects
the semiclassical and quantum spectrum of the $xp$ model to 
form  a continuum. In order to get a discrete spectrum Berry and Keating
introduced the constraints $|x| \geq  \ell_x$ and $|p| \geq \ell_p$, so that
the particle starts at $t=0$ at $(x, p)  = (\ell_x, E/\ell_x)$ and ends 
at $(x, p)  = (E/\ell_p, \ell_p)$ after a time lapse $T= \log (E/\ell_x \ell_p)$
(we assume for simplicity that $x, p>0$).  The trajectories are now bounded,
but  not  periodic. A  semiclassical estimate 
of the number of energy levels, $n_{\rm BK}(E)$, between 0 and $E>0$ is given by the formula 
\beq
n_{\rm BK}(E) =  \frac{A_{\rm BK}}{ 2 \pi \hbar} = 
 \frac{E}{2 \pi \hbar} \left(  \log \frac{ E}{ \ell_x \ell_p} - 1 \right) + \frac{7}{8}, 
\label{cl1}
\eeq
where $A_{\rm BK}$ is the phase space area below the parabola $E = x p$ and the lines
$x = \ell_x$ and $p = \ell_p$, measured in units of the Planck's constant  $2 \pi \hbar$ (see Fig. \ref{xp-curve}). 
The term  $7/8$ arises from the  Maslow phase \cite{BK99}.  In the course of the paper, we shall encounter
this equation several times with the constant term depending on the particular model.

Berry and Keating compared this  result 
 with the average number of Riemann zeros, 
whose imaginary part is less than $t$ with  $t \gg 1$, 

\beq
 \langle n(t) \rangle   \simeq 
 \frac{t}{2 \pi} \left(  \log \frac{ t}{ 2 \pi} - 1 \right) + \frac{7}{8} + O(1/t) \, , 
\label{cl2}
\eeq
finding an agreement  with the identifications 

\beq
t = \frac{ E}{\hbar}, \qquad \ell_x \, \ell_p = 2 \pi \hbar.
\label{cl3}
\eeq
Thus,  the semiclassical energies $E$, expressed in units of $\hbar$,  are  identified
with the Riemann zeros, while $\ell_x \ell_p$ is identified
with the Planck's constant. This result is remarkable
given the simplicity of the assumptions. But one must observe that  the 
derivation of eq.(\ref{cl1}) is heuristic, so one goal is 
to find a consistent quantum version of it. 

Connes proposed another regularization of the $xp$ model  based on the restrictions $|x| \leq  \Lambda$ and $|p| \leq \Lambda$,
 where  $\Lambda$ is a common cutoff, which is taken to infinity at the end of the calculation \cite{C99}. The semiclassical
 number of states is computed as before yielding (see  Fig. \ref{xp-curve}, we set $\hbar =1$)
\beq
n_{\rm C} (E) = \frac{A_{\rm C}}{2 \pi}  =  \frac{E}{2 \pi } \log
  \frac{ \Lambda^2}{ 2 \pi} - 
 \frac{E}{2 \pi}  \left(  \log  \frac{E}{2 \pi}  - 1 \right)  \, . 
\label{11}
\eeq

\noindent 
The first term on the RHS of this formula diverges in the limit  $\Lambda \rightarrow \infty$,
which corresponds to a continuum of states. The second term is minus the
average number of Riemann zeros, which according to Connes, become 
missing spectral lines in the continuum \cite{C99,H76}. This is called the {\em absorption}  spectral interpretation
of the Riemann zeros, as opposed to the standard {\em emission}  spectral interpretation where the {\em zeros}
form a discrete spectrum. Connes, relates  the minus sign in eq.(\ref{11}) 
to a minus sign discrepancy between the fluctuation term of the number 
of zeros and the associated formula in the theory of Quantum Chaos. 
We shall show below  that the negative term in Eq.(\ref{11}) must be seen as 
a finite size correction of discrete  energy  levels and not as an indication of missing spectral lines.  

Let us  give for completeness  the formula  for the exact number of {\em zeros}  up to $t$ \cite{E74,T86}
\barray 
n_{\rm R}(t)  &  = &  \langle n (t) \rangle  + {n}_{\rm  fl}(t),   \label{cl4} \\
\langle n (t) \rangle  & = & 
 \frac{ \theta(t)}{\pi}  +1,   \qquad n_{\rm fl}(t)  =  \frac{1}{ \pi}  {\rm Im} \,  \log \zeta \left( \frac{1}{2} +   i t  \right), 
\nonumber
\earray 
where $\langle n(t) \rangle$ is the  Riemann-von Mangoldt formula that gives the average behavior in terms of the 
function $\theta(t)$
\barray 
\theta(t)  & = &  {\rm Im} \, \log \Gamma \left( \frac{1}{4} +  \frac{i t}{2} \right)  - \frac{t}{2} \log \pi  \; 
\stackrel{t \rightarrow \infty} \longrightarrow  \; 
\frac{t}{2} \log \frac{ t}{2 \pi} - \frac{t}{2} - \frac{ \pi}{8} + O(1/t) \;  , 
\label{cl5} 
\earray 
that can also be written as
\beq
e^{ 2 i \theta(t)} = \pi^{ - i t} \frac{ \Gamma \left( \frac{1}{4} +  \frac{i t}{2} \right) }{ \Gamma \left( \frac{1}{4} -  \frac{i t}{2} \right) }.
\label{cl5b}
\eeq
$\theta(t)$ is the phase of the Riemann zeta function on the critical line, 
that can be expressed as 
\beq
\zeta \left( \frac{1}{2} + i t  \right) = e^{ - i \theta(t)} \, Z(t), \label{cl5c}
\eeq
where $Z(t)$ is the Riemann-Siegel zeta function, or Hardy function, that on the critical line satisfies
\beq
Z(t) = Z(-t) = Z^*(t), \qquad t \in \Rmath \, . 
\label{cl5d}
\eeq

\vspace{0.25 cm} 

{\bf Summary:}  

\vspace{0.25 cm}

\fbox{
\begin{minipage}{42em}
\checkmark \,   The  semiclassical spectrum of the $xp$ Hamiltonian reproduces  the 
 average Riemann zeros. 
  
\xmark  \;  There are two schemes  leading to opposite physical realizations: emission vs absorption.
    
?  \; Quantum version of the semiclassical  $xp$ models.

\end{minipage}
}

\section{The quantum $XP$ model} 

To quantize the  $xp$ Hamiltonian,  Berry and Keating  used   the normal ordered operator  \cite{BK99} 

\beq
\hat{H} =  \frac{1}{2} ( x \, \hat{p}  + \hat{p}  \, x) = - i \hbar \left( x \frac{d}{dx} + \frac{1}{2} \right) , \qquad  x \in \Rmath , 
\qquad 
\label{qu1}
\eeq
where $x$ belongs to the real line and $\hat{p} = - i \hbar d/dx$ is the momentum operator.
We shall show below, that despite of being a natural quantization of the classical $xp$ Hamiltonian, 
it does not reproduce the semiclassical spectrum obtained in the previous section. 
It is however of great interest to study it in detail since it is the basis of the rest of  the work.

It is convenient to restrict $x$ to the  positive half-line, then (\ref{qu1}) is equivalent to the expression
 
\beq
\hat{H} = \sqrt{x} \, \hat{p} \, \sqrt{x},  \qquad  x \geq 0 \,  . 
\qquad 
\label{qu2}
\eeq

$\hat{H}$ is an essentially self-adjoint operator acting  on the Hilbert space $L^2(0, \infty)$ of square integrable
functions in the half line  $ \Rmath_+= (0, \infty)$ \cite{S07a,TM07,ES10}. The eigenfunctions, with eigenvalue $E$,  are given by
\beq
\psi_E(x) = \frac{1}{ \sqrt{2 \pi \hbar}}  \,  x^{- \frac{1}{2} +  \frac{i E}{\hbar} }, \qquad x >  0, \quad   E \in \Rmath \, , 
\label{qu6}
\eeq
 and the  spectrum is the real line $\Rmath$.  The normalization of (\ref{qu6})  is given by  the Dirac's delta
 function 
\beq
\langle \psi_E| \psi_{E'} \rangle
= \int_0^\infty dx \,  \psi_E^*(x) \, \psi_{E'}(x)  = \delta(E - E'). 
\label{qu8}
\eeq
The eigenfunctions (\ref{qu6})  form  an orthonormal basis of $L^2(0, \infty)$,
that  is related to the Mellin transform in the same manner that  the eigenfunctions of the momentum operator $\hat{p}$, 
on the real line,  are related to the Fourier transform \cite{TM07}.
If one takes $x$ in the whole  real line, then  the spectrum of the Hamiltonian  (\ref{qu1})  is   doubly degenerate.
This degeneracy can be understood from the invariance of  $xp$  
under the parity transformation $x \rightarrow -x, p \rightarrow - p$, which
allows one  to split the eigenfunctions with energy $E$ into even and odd sectors
\beq
\psi_E^{(e)}(x) = \frac{1}{ \sqrt{2 \pi \hbar}}  \,  |x|^{- \frac{1}{2} +  \frac{i E}{\hbar} },
\quad \psi_E^{(o)}(x) = \frac{{\rm sign} \,  x}{ \sqrt{2 \pi \hbar}}  \,  |x|^{- \frac{1}{2} +  \frac{i E}{\hbar} },
 \qquad x   \in \Rmath, \quad   E \in \Rmath. 
\label{qu6b}
\eeq
Berry and Keating computed the  Fourier transform of the even wave function 
$\psi_E^{(e)}(x)$  \cite{BK99} 

\barray 
\hat{\psi}_E^{(e)}(p) & = &  \frac{1}{ \sqrt{2 \pi \hbar} } \int_{- \infty}^\infty dx \, \psi_E^{(e)}(x) \, e^{ - i p x/\hbar} 
\label{qu7} \\ 
& = &
\frac{1}{ \sqrt{2 \pi \hbar}}  \,  |p|^{- \frac{1}{2} -  \frac{i E}{\hbar} }  \, (2 \hbar)^{ i E/\hbar} \, 
\frac{ \Gamma \left( \frac{1}{4} +  \frac{i E}{2 \hbar} \right) }{ \Gamma \left( \frac{1}{4} -  \frac{i E}{2 \hbar} \right) } \, , 
\nonumber 
\earray 
which means that the position and momentum eigenfunctions are each other's time reversed,
giving  a physical interpretation of the phase  $\theta(t)$, see  Eq.(\ref{cl5b}). Choosing odd eigenfunctions leads to an  equation
 similar to Eq.(\ref{qu7}) in terms of the gamma functions $\Gamma( \frac{3}{4} \pm \frac{ i E}{2})$ that appear
 in the functional relation of the odd Dirichlet characters.  Equation  (\ref{qu7}) is a consequence of the exchange $x \leftrightarrow p$
symmetry of the $xp$ Hamiltonian, which seems to be an important ingredient of  the  $xp$ model. 

\vspace{0.5 cm}

{\bf Comments:} 

\begin{itemize}

\item  Removing  Connes's cutoff, i.e. $\Lambda \rightarrow \infty$,
gives the quantum Hamiltonians  (\ref{qu1}) or (\ref{qu2}), whose spectrum is a
continuum. This shows that the negative term in Eq.(\ref{11}) does
not correspond  to missing spectral lines. In the next section we give a physical
interpretation of this term in another context.

\item $xp$  is  invariant under the 
scale transformation (dilations)  $x \rightarrow K  x, p \rightarrow K^{-1}  p$,  with $K >0$. 
An example of this transformation is 
the classical trajectory (\ref{cl0}), whose infinitesimal generator is  $xp$. 
Under dilations,  $\ell_x \rightarrow K  \ell_x,  \; \ell_p \rightarrow K^{-1}  \ell_p$, 
so  the  condition  $\ell_x \ell_p = 2 \pi \hbar$ is preserved.
Berry and Keating suggested to use integer dilations $K =n$, corresponding to
evolution times $\log n$, to write  \cite{BK99} 
\beq
\psi_E(x) \rightarrow \sum_{n=1}^\infty \psi_E(nx) = \frac{1}{ \sqrt{2 \pi \hbar}}  \,  x^{- \frac{1}{2}+  \frac{i E}{\hbar}}  
\sum_{n=1}^\infty \frac{1}{n ^{ \frac{1}{2} -   \frac{i E}{\hbar}}}  = \frac{1}{ \sqrt{2 \pi \hbar}}  \,  x^{- \frac{1}{2}+  \frac{i E}{\hbar}}  \zeta( 1/2 -  i E/\hbar) \, . 
\label{psimod}
\eeq
If  there  exists  a physical reason for this quantity to vanish one would obtain the Riemann zeros $E_n$.
Equation (\ref{psimod}) could be interpreted as the breaking of the continuous scale invariance to discrete scale invariance.


\end{itemize}

\vspace{0.25 cm} 

{\bf Summary:}  

\vspace{0.25 cm}

\fbox{
\begin{minipage}{42em}

\xmark \,   The  normal order quantization of $xp$  does not exhibit  any  trace  of the Riemann zeros. 
  
\checkmark  \; The phase of the zeta function appears in the Fourier transform of the  $xp$ eigenfunctions.

\end{minipage}
}

\section{The Landau model and $XP$}

Let us consider a charged particle moving in a plane under the action of a perpendicular
magnetic field and an electrostatic potential  $V(x,y) \propto xy$  \cite{ST08}. The Langrangian
describing the dynamics is given, in the Landau gauge, by
\beq
{\cal L}  =  \frac{\mu}{2} ( \dot{x}^2 + \dot{y}^2 ) - \frac{e B }{c} \dot{y} x - e \lambda x y  \, , 
\label{L12}
\eeq
\noindent
where $\mu$ is the mass, $e$ the electric charge, $B$ the magnetic field, $c$ the speed
of light and $\lambda$ a coupling constant  that parameterizes  the electrostatic potential. There are
two normal modes with  real,  $\omega_c$, and imaginary, $\omega_h$, angular
frequencies, describing  a cyclotronic and a hyperbolic motion  respectively. In the limit
where $\omega_c >> |\omega_h|$, only the Lowest Landau Level (LLL)  is relevant and the effective Lagrangian becomes
\beq
{\cal L}_{\rm eff}   =  p \dot{x} - |\omega_h |  x p, \qquad p = \frac{\hbar y}{\ell^2}, \qquad
\ell = \left(  \frac{ \hbar c}{ e B} \right)^{1/2},
\label{L13}
\eeq
\noindent  
where $ \ell$  is the magnetic length, which is proportional to 
the radius of the cyclotronic orbits in the LLL.  The coordinates $x$ and $y$, which
commute in the 2D model, after the proyection to the LLL become canonical
conjugate variables,  and the effective Hamiltonian is proportional to the 
 $xp$ Hamiltonian with the proportionality constant given by the angular frequency $|\omega_h|$ 
 (this is the missing frequency factor mentioned in section II). 
The quantum Hamiltonian associated to the Lagrangian (\ref{L12}) is
\beq
\hat{H} = \frac{ 1}{2 \mu} \left[ \hat{p}_x + \left( \hat{p}_y + \frac{ \hbar}{ \ell^2} x \right)^2 \right]   + e \lambda x y \, , 
\label{L14}
\eeq
where  $\hat{p}_x = - i \hbar \partial_x$ and $\hat{p}_y = - i \hbar \partial_y$. After a unitary transformation (\ref{L14})  becomes
the  sum of two commuting Hamiltonians corresponding to the cyclotronic and hyperbolic motions alluded  to  above 
\barray 
H  & = &  H_c + H_h, \label{L16} \\
H_c &  = &  \frac{ \omega_c}{2} ( \hat{p}^2 + q^2 ),    \quad  
\qquad H_h  =  \frac{ |\omega_h|}{2} ( \hat{P} Q + Q \hat{P} ). 
\nonumber 
\earray 
In the limit $\omega_c \gg |\omega_h|$ one has 
\beq
\omega_c  \simeq \frac{ e B}{\mu c},  \quad  |\omega_h| \sim \frac{ \lambda c}{B}  \, . 
\label{L17}
\eeq
The unitary transformation that brings  Eq. (\ref{L14}) into Eq.(\ref{L16}) corresponds
to the classical canonical transformation

\beq
q= x + p_y , \quad p = p_x, \quad Q = - p_y, \quad P = y + p_x \, . 
\label{L18}
\eeq
When  $\omega_c \gg |\omega_h|$, the low energy states of $H$ are the product of the
lowest eigenstate of $H_c$, namely  $\psi= e^{ - q^2/2 \ell^2}$, times  the eigenstates of $H_h$
that can be chosen as even or odd under the parity transformation $Q \rightarrow - Q$

\beq
\Phi^+_E(Q) = \frac{1}{ |Q|^{ \frac{1}{2} - i E}},  \qquad \Phi^-_E(Q) = \frac{{\rm sign} (Q)}{ |Q|^{ \frac{1}{2} - i E}} \, . 
\label{L19}
\eeq
The corresponding wave functions are given by  (we choose $|\omega_h| =1$) 

\beq
\psi_E^\pm(x,y) = C \int dQ  \;  e^{- i Q y/\ell^2} \, e^{ - (x-Q)^2/2 \ell^2} \Phi^\pm_E(Q)  \, , 
\label{L20}
\eeq 
where $C$ is a normalization constant, which yields

\begin{figure}[h!]
\begin{center}
\includegraphics[width=.35\linewidth]{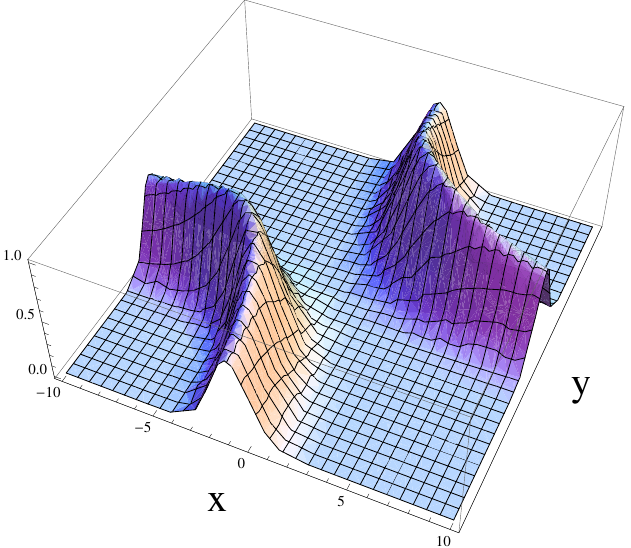}
\hspace{1.5cm} 
\includegraphics[width=.3\linewidth]{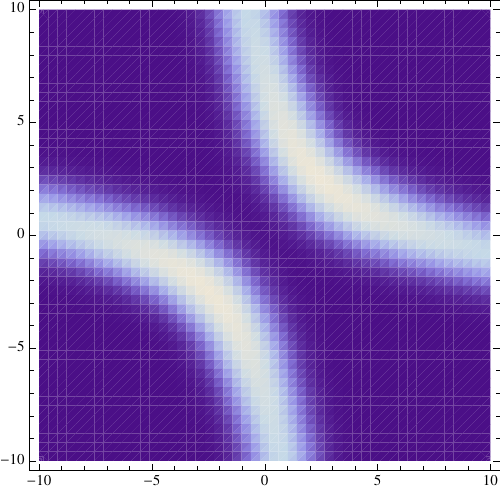}
\end{center}
\caption{Plot of $|\psi_E^+(x,y)|$ for $E=10$ in the region $- 10 < x, y < 10$. Left: 3D representation,
Right: density plot. 
  }
\label{psi-landau}
\end{figure}

\barray 
\psi_E^+(x,y) & = & C_E^+  e^{ -  \frac{ x^2}{ 2 \ell^2} } \, M\left( \frac{ 1}{4} + \frac{ i E}{2}, \frac{ 1}{2}, \frac{ (x-iy)^2}{2 \ell^2} \right) \, ,  \label{L17} \\
\psi_E^-(x,y) & = & C_E ^-  (x- i y) e^{ -  \frac{ x^2}{ 2 \ell^2} } \, M\left( \frac{ 3}{4} + \frac{ i E}{2}, \frac{ 3}{2}, \frac{ (x-iy)^2}{2 \ell^2} \right)  \, , \nonumber 
\earray 
where $M(a,b,z)$ is a confluent hypergeometric function \cite{AS72}. Fig.\ref{psi-landau}
shows that the maximum of the absolute value of $\psi^+_E$  is attained on the 
classical trajectory $E = x y$ (in units of $\hbar = \ell=1$). This 2D  representation 
of the classical trajectories is possible because in the LLL $x$ and $y$ become canonical
conjugate variables and consequently  the 2D plane coincides with the phase space $(x,p)$. 

To count the number of states with an energy below $E$ one places the particle 
into  a box: $ |x| < L, |y| < L$ and impose the boundary conditions  
\beq
\psi^+_E(x, L) =  e^{i x L/\ell^2} \, \psi^+_E(L,x) \, , 
\label{L18}
\eeq
which identifies the outgoing particle at $x=L$ with the incoming particle at $y=L$
up to a phase. The asymptotic behavior $L \gg \ell$  of (\ref{L17}) is
\barray 
\psi_E^+(L,x)  & \simeq &  e^{- i x L/\ell^2 - x^2/2 \ell^2} \frac{ \Gamma \left( \frac{1}{2} \right) }{ \Gamma \left( \frac{1}{4} + \frac{ i E}{2} \right)} 
\left( \frac{ L^2}{ 2 \ell^2} \right)^{- \frac{1}{4} + \frac{ i E}{2}}  \, , 
\label{L19}  \\
\psi_E^+(x,L)  & \simeq &   e^{- x^2/2 \ell^2} \frac{ \Gamma \left( \frac{1}{2} \right) }{ \Gamma \left( \frac{1}{4} -  \frac{ i E}{2} \right)} 
\left( \frac{ L^2}{ 2 \ell^2} \right)^{- \frac{1}{4} -  \frac{ i E}{2}}  \, ,  \nonumber 
\earray 
that  plugged  into the BC (\ref{L18}) yields 
\beq
\frac{  \Gamma \left( \frac{1}{4} + \frac{ i E}{2} \right)}{ \Gamma \left( \frac{1}{4} -  \frac{ i E}{2} \right)}
\left( \frac{ L^2}{ 2 \ell^2} \right)^{ -  i E}  = 1 \, , 
\label{L20}
\eeq
or using Eq.(\ref{cl5b})
\beq
e^{ 2 i \theta(E) } 
\left( \frac{ L^2}{ 2 \pi \ell^2} \right)^{ -  i E}  = 1 \, . 
\label{L201}
\eeq
Hence  the number of states $n(E)$ with energy less that $E$ is given by 
\beq
n(E)  \simeq \frac{E}{2 \pi} \log \left( \frac{ L^2}{ 2 \pi \ell^2} \right) + 1 - \langle n(E) \rangle  \, , 
\label{L21}
\eeq
whose asymptotic behavior  coincides with Connes's formula  (\ref{11})
for a cutoff $\Lambda = L/\ell$. In fact,  the term $\langle n(E) \rangle$ is the  exact 
Riemann-von Mangoldt formula (\ref{cl4}).

\vspace{0.25cm}




%


\vspace{0.25 cm} 

{\bf Summary:} 

\vspace{0.25 cm}

\fbox{
\begin{minipage}{49em}
\checkmark The Landau model with a $xy$ potential provides a physical realization of Connes's $xp$ model. 

    
\checkmark   The finite size effects in the spectrum are given by the Riemann-von Mangoldt  formula. 

\xmark \;  There are no missing spectral lines in the physical  realizations of $xp$ \`a la Connes.

\end{minipage}
}

\section{The  $XP$ model revisited} 

\begin{figure}[t!]
\begin{center}
\includegraphics[width=.30\linewidth]{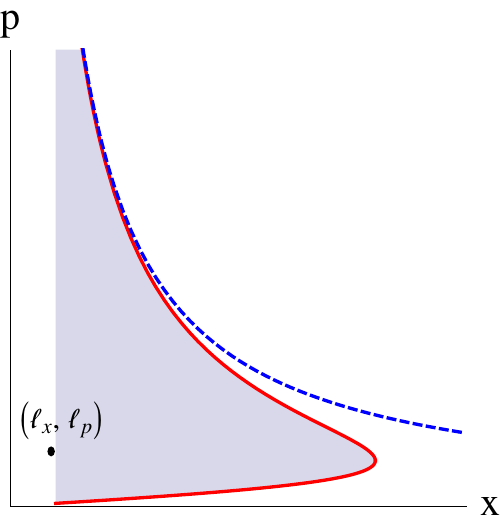}
\hspace{0.4cm}
\includegraphics[width=.30\linewidth]{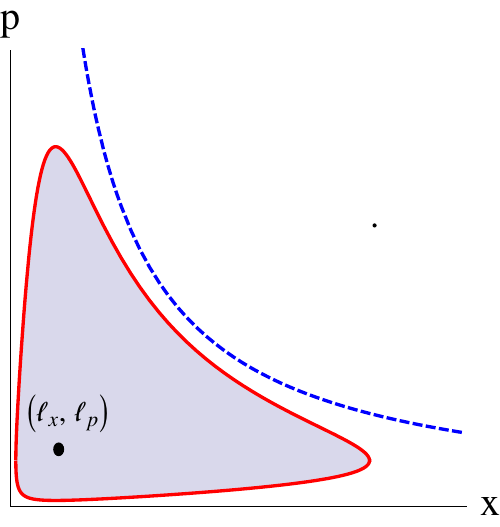}
\end{center}
\caption{
Classical trajectories  of the Hamiltonians (\ref{q1}) (left) and (\ref{q1b}) (right)  in phase space with $E>0$. 
The dashed lines denote  the hyperbola $E = xp$. 
$(\ell_x, \ell_p)$ is a fixed  point solution of the classical equations generated by  (\ref{q1}) and (\ref{q1b}).
  }
\label{tray}
\end{figure}

An intuitive argument of why the quantum Hamiltonian $(x \hat{p} + \hat{p}x)/2$ has a continuum
spectrum is that  the classical trajectories of $xp$ are unbounded. Therefore,  to have 
a discrete spectrum one should modify $xp$ in order to bound the trajectories. 
This  is achieved by the classical Hamiltonian  \cite{SL11} 
\beq
{H}_{ I} = x \left(  p + \frac{ \ell_p^2}{p} \right),  \quad x \geq \ell_x.
\label{q1}
\eeq
For   $|p| >> \ell_p$,  a classical trajectory with energy $E$ satisfies  $E \simeq x p$, 
but for $|p| \sim  \ell_p$, the coordinate  $|x|$  slows down,
reaches a maximum and goes back to the value  $\ell_x$, where it bounces
off starting again at high momentum. In this manner one gets a periodic orbit 
(see   fig. \ref{tray})  

\barray 
x(t) &= & \frac{ \ell_x}{ |p_{0}|}  e^{ 2 t} \sqrt{  ( p_0^2 + \ell_p^2) e^{- 2 t} - \ell_p^2}, \qquad 0 \leq t \leq T_E  \, , 
\nonumber  \\
p(t) & =  & \pm \sqrt{ ( p_0^2 + \ell_p^2)  e^{ - 2 t} - \ell_p^2} \, , 
\label{q5} 
\earray 
where  $T_E$ is the period given by  (we take $E>0$)

\beq
T_E = \cosh^{-1}   \frac{E}{2 {\ell_x \ell_p}} \rightarrow \log \frac{E}{{\ell_x \ell_p}} \quad  (E \gg  {\ell_x \ell_p})  \, . 
\label{q6}
\eeq
The asymptotic value  of $T_E$ is the time lapse  it takes 
a particle to go from  $x= \ell_x$ to   $x = E/\ell_p$ in the $xp$ model. 

The exchange symmetry $x \leftrightarrow p$  of $xp$ is broken by the Hamiltonian (\ref{q1}).
To restore it,  Berry and Keating proposed the $x-p$ symmetric   Hamiltonian  \cite{BK11}

\beq
{H}_{II} = \left(  x + \frac{ \ell_x^2}{x} \right) \left(  p + \frac{ \ell_p^2}{p} \right),   \quad x \geq 0.
\label{q1b}
\eeq
Here  the classical trajectories  turn clockwise around the point $(\ell_x, \ell_p)$, 
and for $x \gg \ell_x$ and $p \gg  \ell_p$, approach the parabola $E = xp$  
(see  Fig. \ref{tray}).  The semiclassical analysis of (\ref{q1}) and (\ref{q1b})
reproduce  the  asymptotic behavior of Eq.(\ref{cl1}) to leading orders $E \log E$ and $E$, 
but differ in the remaining terms. 

\begin{figure}[t!]
\begin{center}
\includegraphics[width=.35 \linewidth]{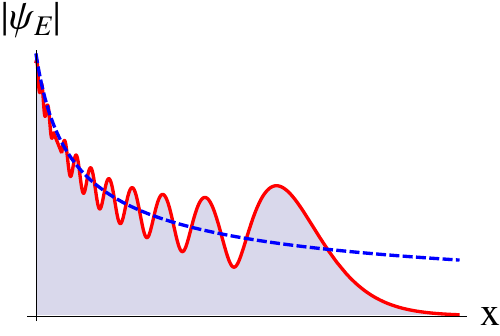}
\end{center}
\caption{
Absolute values of the wave function  $\psi_E(x)$, given in eq.(\ref{h6}) (continuous line),  and 
$x^{- \frac{1}{2}}$ (dashed  line). }
\label{psi2}
\end{figure}

The two models discussed above have the general form 
\beq
{H}_{} = U(x) p + \ell_p^2  \frac{ V(x)}{p}, \quad x \in D, 
\label{h1}
\eeq
where $U(x)$ and $V(x)$ are positive functions defined  in an interval  $D$ of the real line. 
$H_{I}$ corresponds to $U(x) = V(x) = x, D= (\ell_x, \infty)$,
and  $H_{II}$ corresponds to $U(x) = V(x) = x + \ell_x^2/x, D= (0, \infty)$. 
The classical Hamiltonian (\ref{h1}) can be  quantized  in terms of the operator 
\beq
\hat{H} = \sqrt{U} \,  \hat{p} \,   \sqrt{U} + \ell^2_p \,  \sqrt{V} \,  \hat{p}^{-1}  \,   \sqrt{V},
\label{h2}
\eeq
where  $\hat{p}^{-1}$ is pseudo-differential operator 

\beq
\left( \hat{p}^{-1} \psi \right)(x) = - \frac{i}{ \hbar} \int_x^\infty dy \,  \psi(y), 
\label{h3}
\eeq
which satisfies that $\hat{p}  \, \hat{p}^{-1} =  \hat{p}^{-1}   \hat{p} = {\bf 1}$ acting on functions which vanish
in the limit  $x \rightarrow \infty$.  The action of $\hat{H}$  is 
\barray 
( \hat{H} \psi)(x) &  = &  - i  \hbar\sqrt{U(x)}  
 \frac{ d}{dx}  \left\{  \sqrt{U(x)}   \psi(x))   \right\}   -  \frac{ i \ell_p^2}{ \hbar} \int_x^\infty dy \, \sqrt{V(x) V(y)}  \, \psi(y). 
   \label{h4} 
\earray 
The normal order prescription that leads from  (\ref{h1}) to  (\ref{h4}) will be derived in section VII  in the case where $U(x) = V(x) =x$, 
but holds in general \cite{MS12}.  We want the Hamiltonian (\ref{h2}) to be self-adjoint, that is \cite{vN,GP90}
\beq
\langle \psi_1 | \hat{H} |\psi_2 \rangle = \langle  \hat{H} \psi_1   |\psi_2 \rangle \, . 
\label{h4b}
\eeq
When  the interval is   $D = (\ell_x, \infty)$,  Eq.(\ref{h4b}) holds  for   wave functions  that vanishes sufficiently
fast at infinity and  satisfy   the  non  local boundary condition
\beq
\hbar  \,  e^{ i \vartheta} \, \sqrt{U(\ell_x)} \,  \psi(\ell_x)  =   \ell_p \int_{\ell_x}^\infty  dx \, \sqrt{V(x)}  \, \psi(x), 
\label{h5}
\eeq
where $\vartheta \in [0, 2 \pi)$ parameterizes the self-adjoint extensions of $\hat{H}$.
The quantum Hamiltonian associated to  (\ref{q1}) is  

\beq
\hat{H}_{I} = \sqrt{x} \, \hat{p} \, \sqrt{x} + \ell_p^2 \, \sqrt{x} \, \hat{p}^{-1}  \, \sqrt{x}, \qquad x \geq \ell_x \, , 
\label{5b}
\eeq 
and its  eigenfunctions  are proportional to  (see fig. \ref{psi2})
 
\barray 
\psi_E(x)  &  = &  x^{ \frac{i E}{ 2 \hbar}} \, K_{\frac{1}{2} - \frac{ i E}{2 \hbar}} \left( \frac{ \ell_p x}{ \hbar} \right) 
\propto  \left\{
\begin{array}{ll}
x^{ - \frac{1}{2} + \frac{ i E}{\hbar}} & x \ll \frac{E}{ 2 \ell_p} ,  \\ 
x^{ - \frac{1}{2} + \frac{ i E}{2 \hbar}}  e^{ - \ell_p x/\hbar}  & x \gg \frac{E}{ 2 \ell_p}  ,  \\
\end{array}
\right.
\label{h6} 
\earray 
where $K_\nu(z)$ is the modified $K$-Bessel function \cite{AS72}. For small values of $x$, 
 the wave functions (\ref{h6})  behave  as those of the $xp$ Hamiltonian,  given in Eq.(\ref{qu6}),
while  for large values of $x$ they  decay  exponentially giving a normalizable state.
The boundary condition (\ref{h5}) reads in this case
\beq
\hbar  \,  e^{ i \vartheta} \, \sqrt{ \ell_x} \,  \psi(\ell_x)  =   \ell_p \int_{\ell_x}^\infty  dx \, \sqrt{x}  \, \psi(x), 
\label{h555}
\eeq
and substituting (\ref{h6})  yields  the equation for  the eigenenergies $E_n$, 
\beq 
e^{  i \vartheta} K_{\frac{1}{2} - \frac{ i E}{2 \hbar}} \left( \frac{  \ell_x \ell_p}{ \hbar} \right) -  
 K_{\frac{1}{2} +  \frac{ i E}{2 \hbar}} \left( \frac{  \ell_x \ell_p}{ \hbar} \right) = 0.
\label{h7}
\eeq
For  $\vartheta =0 $ or $\pi$, the eigenenergies form time reversed pairs $\{ E_n, - E_n \}$, and  for  $\vartheta = 0$,
 there is a zero energy  state $E=0$. Considering  that 
the Riemann zeros form  pairs   $s_n= 1/2 \pm i t_n$, 
with $t_n$ real under the RH, and that  $s=1/2$  is not a  {\em zero}  of $\zeta(s)$, we are led to the choice
 $\vartheta =\pi$.  On the other hand,   using the  asymptotic behavior 
\barray 
K_{a +  \frac{i t}{2}}(z)  & \longrightarrow  &  \sqrt{  \frac{\pi}{t} }    \left( \frac{t}{z} \right)^{a} \, e^{ - \pi t/4}  \; e^{  \frac{i \pi}{2} ( a- \frac{1}{2} ) }  \left(
\frac{t}{ z e} \right)^{ i t/2}, \qquad a >0, t \gg 1, 
\label{h8} 
\earray 
one   derives  in the limit $|E| \gg \hbar$, 
\beq
 K_{\frac{1}{2} +  \frac{ i E}{2 \hbar}} \left( \frac{  \ell_x \ell_p}{ \hbar} \right) 
+ K_{\frac{1}{2} - \frac{ i E}{2 \hbar}} \left( \frac{  \ell_x \ell_p}{ \hbar} \right)  = 0  \longrightarrow
 \cos \left( \frac{ E}{2 \hbar} \log \frac{ E}{ \ell_x \ell_p e} \right) = 0  \, , 
\label{h9}
\eeq
hence  the number of eigenenergies in the interval $(0,E)$ is  given asymptotically  by 
\beq
 n(E)  \simeq \frac{ E}{ 2  \pi \hbar } \left(  \log \frac{E}{  \ell_x \ell_p } - 1 \right) - \frac{1}{2}   + O(E^{-1}).
\label{h10}
\eeq
This equation   agrees with the leading terms of the semiclassical spectrum (\ref{cl1})
and the average Riemann zeros (\ref{cl2}) under the identifications (\ref{cl3}).   
Concerning  the classical  Hamiltonian  (\ref{q1b}),  Berry and Keating obtained, by a 
 semiclassical analysis, the  asymptotic behavior of the counting function $n(E)$
 \beq
 n_{}(t)  \simeq 
 \frac{t}{2 \pi} \left(  \log \frac{ t}{ 2 \pi} - 1 \right) - \frac{ 8 \pi}{t}  \log \frac{ t}{ 2 \pi}  + \dots,  \quad t  \gg 1,  
\label{bk1}
\eeq
 where $t = E/\hbar$ and $\ell_x \ell_p = 2 \pi \hbar$. Again,  the first two leading terms agree with 
 iemann's formula (\ref{cl2}), while  the next leading corrections are different from  (\ref{h10}). 
In both cases,  the   constant $7/8$ in  Riemann's formula (\ref{cl2}) is missing. 

\vspace{0.25 cm}

{\bf Summary:} 

\vspace{0.25 cm}

\fbox{
\begin{minipage}{44em}

\checkmark The  Berry-Keating $xp$  model can be implemented quantum mechanically. 
  
\xmark  \;  The classical $xp$ Hamiltonian has to be modified with ad-hoc  terms  to have bounded trajectories.

\xmark \; In the quantum theory the latter terms become non local operators. 
  
\xmark \; The modified   $xp$ quantum Hamiltonian  related to the average  Riemann zeros  is  not unique. 

\xmark \;  There is no trace of the exact Riemann zeros in the spectrum of the modified $xp$ models. 

\end{minipage}
}

\section{The space-time geometry of the  modified $XP$ models}

In this section we show that the  modified $xp$ Hamiltonian (\ref{h1}) 
is  a disguised general theory of relativity \cite{S12}. 
Let us  first  consider  the Langrangian of the $xp$ model,
\beq
L = p \dot{x} - H = p   \dot{x}  -  xp    \, . 
\label{g0}
\eeq
In classical mechanics,  where $H = p^2/2m + V(x)$,  the Lagrangian can be expressed solely 
in terms of  the position $x$ and velocity $\dot{x}= dx/dt$. This is achieved by writing
the momentum in terms of the velocity by means of the Hamilton  equation  
 $\dot{x} = \partial H/\partial p =p/m$.  However, in  the  $xp$ model  the momentum $p$
is not a function of  the velocity because $\dot{x} = \partial H/\partial p = x$. Hence the Lagrangian
(\ref{g0}) cannot be expressed uniquely  in terms of $x$ and $\dot{x}$. 
The situation changes radically  for the Hamiltonian (\ref{h1})
whose Lagrangian is  given by 
\beq
L = p \, \dot{x}  -  H  = p \, \dot{x} -  U(x) p -  \ell_p^2  \frac{ V(x)}{p}. 
\label{g1}
\eeq
Here the  equation of motion 
\beq
\dot{x} = \frac{ \partial H}{ \partial p} = U(x) -  \ell_p^2  \frac{ V(x)}{p^2}, 
\label{g1b}
\eeq
allows one  to  write   $p$ in terms of $x$ and $\dot{x}$,  
\beq
p =  \eta \ell_p  \sqrt{ \frac{V(x)}{ U(x) - \dot{x}}}  \, , \qquad \eta = {\rm sign} \, p  \, , 
\label{g2}
\eeq
where  $\eta = \pm 1$ is the sign of the momentum that  is a conserved quantity. 
The positivity of $U(x)$ and $V(x)$, imply 
that the velocity $\dot{x}$ must never exceed  the value of $U(x)$.  
Substituting (\ref{g2}) back into (\ref{g1}),  yields  the action 
\beq
S_\eta  =   -   \ell_p \eta  \int \sqrt{ - ds^2  } \, , 
\label{g6}
\eeq
which, for either  sign of $\eta$,  is the action of a relativistic particle moving in a 1+1 dimensional
spacetime metric
\beq 
ds^2 = 4 V(x) ( - U(x)  dt^2 + dt dx )  \, . 
\label{g6b}
\eeq
The parameter  $\ell_p$ plays the role of $mc$ where $m$ is the mass of the particle and $c$
is the speed of light. This result implies  that the classical trajectories of the Hamiltonian
(\ref{h1}) are the geodesics of the metric (\ref{g6b}). The unfamiliar form of (\ref{h1}) is 
due to a special choice of spacetime coordinates where the component $g_{xx}$ of the metric 
vanishes. A diffeomorphism of  $x$ permits to set  $V(x) = U(x)$. 
The scalar curvature of the metric (\ref{g6b}),  in this {\em gauge},  is 

\beq
R(x) = - 2 \frac{  \partial_x^2 V(x)}{ V(x)} \, , 
\label{g7}
\eeq
and vanishes for the models  $V(x) = x$ and $V(x) = {\rm constant}$.
For the Hamiltonian (\ref{q1b}) one obtains  $R(x) = - 4 \ell_x^2/( x ( x^2 + \ell_x^2))$ 
which  vanishes asymptotically. 

\begin{figure}[t]
\begin{center}
\includegraphics[height= 6.0 cm]{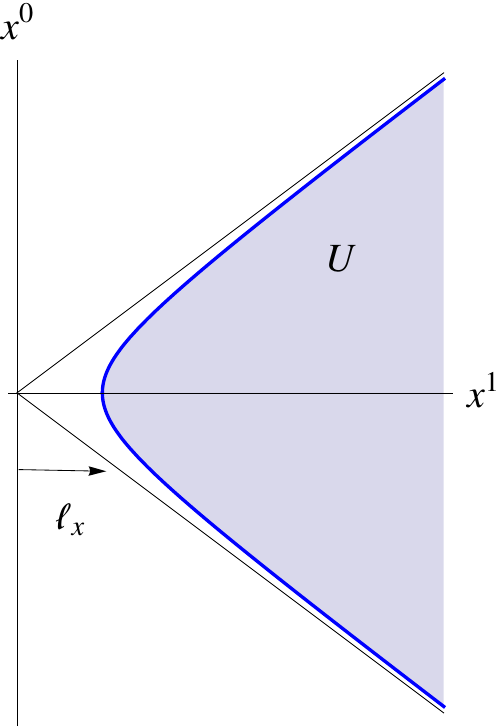} \hspace{1 cm}
\includegraphics[height= 6.0 cm]{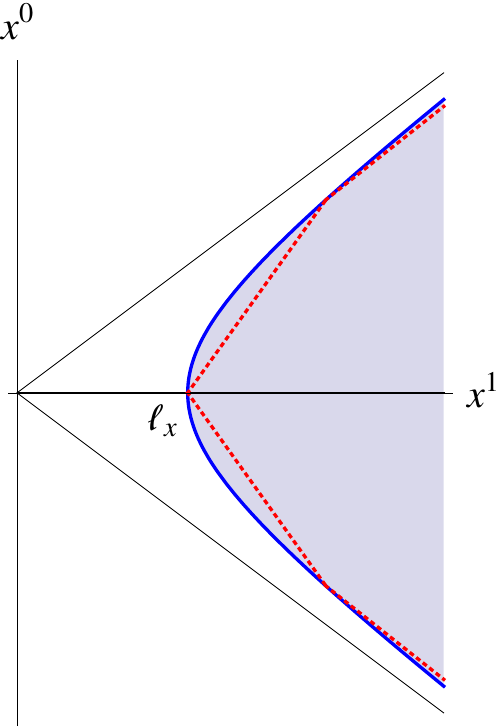}
\end{center}
\caption{ Left: Domain ${\cal U}$  of  Minkowski spacetime given in Eq.(\ref{uspace}). Right: 
The  classical trajectory  given in Eq.(\ref{q5}), and  plotted in Fig.\ref{tray}-left,  becomes a  straighline 
that bounces  off regularly at  the  boundary (dotted line). 
}
\label{U-space}
\end{figure}

The flatness of the  metric associated to  the Hamiltonian (\ref{q1}) 
 implies the existence of  coordinates  $x^0, x^1$ 
where (\ref{g6b})  takes the   Minkowski  form 
 
 \beq
ds^2  =   \eta_{\mu \nu} dx^\mu dx^\nu,  \qquad {\rm diag} \; \eta_{ \mu \nu} = (-1, 1) \, . 
\label{41d}
\eeq
 The change of variables is given by 
\beq
t = \frac{1}{2} \log ( x^0 + x^1), \qquad  x =  \sqrt{ -(x^0)^2 + (x^1)^2}  \, . 
\label{11b}
\eeq
Let  ${\cal U}$ denote the space-time  domain of the   model. In both  coordinates it reads  
 \barray 
{\cal U}  & =  &  \left\{ (t,x) \;   | \;  t \in (- \infty, \infty), \;    x \geq \ell_x  \right\}  =
  \left\{ (x^0,x^1) \;   | \;  x^0 \in (- \infty, \infty), \;    x^1 \geq \sqrt{(x^0)^2 + \ell_x^2}  \right\}.
\label{uspace}
\earray 
The  boundary of  ${\cal U}$,  denoted by $\partial {\cal U}$,  is the   hyperbola $x^1  =  \sqrt{(x^0)^2 + \ell_x^2}$, 
that passes  through the point $(x^0, x^1)= (0, \ell_x)$, 
 (see fig \ref{U-space}).  
A convenient parametrization of the coordinates $x^{\mu}$  is given by the 
Rindler variables  $\rho$ and $\phi$ \cite{R66}

\beq
x^0 = \rho \, \sinh \phi, \qquad x^1 = \rho \, \cosh \phi, 
\label{ri1}
\eeq
or in light-cone coordinates
\beq
x^\pm  =  x^0 \pm x^1 = \pm \rho e^{ \pm \phi} , 
\label{ri1b}
\eeq
where the Minkowski   metric becomes 
\beq
ds^2 = - dx^+ \,  dx^- =  d \rho^2 -  \rho^2 d \phi^2 \, . 
\label{r1c}
\eeq
These coordinates describe the right wedge  of Rindler spacetime in 1+1 dimensions
\barray 
{\cal R}_+   & = &    \left\{ (x^0,x^1) \;   | \;  x^0 \in (- \infty, \infty), \;    x^1 \geq |x^0|  \right\}  
=    \left\{ (\rho, \phi) \;   | \;  \phi  \in (- \infty, \infty), \;   \rho > 0   \right\}.
\label{ri2} 
\earray 
Notice that ${\cal U} \subset {\cal R}_+$. The boundary $\partial {\cal U}$ corresponds to 
the hyperbola $\rho = \ell_x$ that  is the worldline of a particle  moving with uniform
 acceleration equal to $1/\ell_x$ (in units $c=1$). The Rindler variables are the 
 ones used to  study the Unruh effect  \cite{U76}.

Let us now consider the classical Hamiltonian (\ref{q1b}). The underlying metric  is given by
Eq.(\ref{g6b}) with $U(x) = V(x) = x + \ell_x^2/x$.  The change of variables

\beq
t = \frac{1}{2} \log ( x^0 + x^1), \qquad x =  \sqrt{ -(x^0)^2 + (x^1)^2 - \ell_x^2} \, , 
\label{ri4}
\eeq
brings the metric to the form 
\beq
ds^2 = \frac{ -(x^0)^2 + (x^1)^2 }{ -(x^0)^2 + (x^1)^2 - \ell_x^2} \,  \eta_{\mu \nu} dx^\mu dx^\nu = 
\frac{\rho^2}{ \rho^2 - \ell_x^2} ( d \rho^2 - \rho^2 d \phi^2), \qquad \rho \geq \ell_x  \, , 
\label{ri5}
\eeq
which in the limit $\rho \rightarrow \infty$ converges to the flat metric (\ref{r1c}).

\vspace{0.25 cm}

{\bf Summary: } 

\vspace{0.25 cm}

\fbox{
\begin{minipage}{46em}

\checkmark The classical modified $xp$ models  are general relativistic theories in 1+1 dimensions.

\checkmark $H = x (p + \ell_p^2/p)$ is related to a  domain ${\cal U}$  of Rindler   space-time. 

\checkmark $l_p$ is the mass of the particle. 

\checkmark  $1/\ell_x$ is  the acceleration of  a particle whose worldline is the boundary of  ${\cal U}$. 

? \, Relativistic quantum  field theory  of the modified $xp$ models. 

\end{minipage}
}

\section{Diracization of   $H = X (P + \ell_p^2/P)$}

In this section we show that the Dirac theory  provides the relativistic
quantum version of the modified $xp$ models \cite{S14}. We shall focus
on the classical Hamiltonian $H =  x (p + \ell_p^2/p)$ because the
flatness of the associated  space-time makes the computations easier, but the result is general:
the quantum Hamiltonian (\ref{h4}) can be derived  from the Dirac equation in  a curved space-time with metric (\ref{g6b}) \cite{MS12}.

The Dirac action of  a fermion with  mass $m$   in the spacetime domain (\ref{uspace})
is given by  (in units $\hbar = c=1$)  
\beq
S =  \frac{i}{2} \int_{\cal U} dx^0 dx^1 \, \bar{\psi}  ( {\slashed \partial} + i m)  \psi  \, , 
\label{d0}
\eeq
 where  $\psi$ is a two component spinor,  $\bar{\psi} = \psi^\dagger \gamma^0$, 
  ${\slashed \partial}  = \gamma^\mu \partial_\mu \; (\partial_\mu = \partial/ \partial x^\mu)$,
 and $\gamma^\mu$ are  the 2d Dirac matrices written in terms of the Pauli matrices $\sigma^{x,y}$ as 
\beq
\gamma^0 = \sigma^x, \qquad \gamma^1 = - i \sigma^y, \qquad
\psi = \left( 
\begin{array}{c}
\psi_- \\
\psi_ + \\
\end{array}
\right)  \, .
\label{d2}
\eeq
The variational principle applied to (\ref{d0}) provides   the Dirac equation 
\beq
( {\slashed \partial}  + i  m) \psi  = 0  \, , 
\label{d1}
\eeq
and the boundary condition
\barray 
 \dot{x}^- \psi_-^\dagger  \delta \psi_-  -  \dot{x}^+ \psi_+^\dagger   \delta \psi_+ = 0, 
\label{d1b}
\earray 
where  $\dot{x}^\pm =  d x^\pm/d \phi=   \ell_x e^{ \pm \phi}$ is the vector tangent to   the boundary $ \partial {\cal U}$
in the light-cone coordinates $x^\pm = x^0 \pm x^1$.  The Dirac equation reads in components 
\barray
( \partial_0 -   \partial_1 ) \psi_+ + i  m \psi_-  & = & 0, \qquad  ( \partial_0  +    \, \partial_1 ) \psi_-  + i   m  \psi_+ = 0 \, . 
\label{d3}
\earray 
If $m=0$ then $\psi_\pm$  depends only $x^\pm$, and so the  fields  propagate to the left,  $\psi_+(x^+)$,    or to the right,  $\psi_-(x^-)$,   at the speed of light. 
The derivatives in Eq.(\ref{d3}) can be written in terms 
the variables $t$ and $x$  using   Eq.(\ref{11b}), 
\beq
\partial_0 - \partial_1 = -  \frac{2 e^{2 t}}{ x} \partial_x, 
\qquad \partial_0 + \partial_1 = e^{ - 2 t}  ( \partial_t + x \partial_x ) \, . 
\label{d4}
\eeq 
Let us denote by  $\tilde{\psi}_\mp(t,x)$ the fermion fields in the coordinates $t,x$
and by $\psi_\mp(x^0, x^1)$ the fields in the coordinates $x^0, x^1$. The relation between these  fields is given by 
the transformation law
\beq
\psi_- = \left( \frac{  \partial x }{ \partial x^-  }  \right)^{  \frac{1}{2} } \tilde{\psi}_-  = ( 2 x)^{- \frac{1}{2}}  \,  e^t \tilde{\psi}_- , \quad
\psi_+  = \left( \frac{  \partial x }{ \partial  x^+ }  \right)^{ \frac{1}{2} } \tilde{\psi}_-  = (x/2)^{ \frac{1}{2} }   \,  e^{-t}  \tilde{\psi}_+  \, . 
\label{d5}
\eeq
Plugging Eqs.(\ref{d4}) and (\ref{d5}) into (\ref{d3})  gives 

\beq
i \partial_t \,  \tilde{\psi}_- = - i \sqrt{x} \partial_x  \left( \sqrt{x} \tilde{\psi}_- \right) + m x \tilde{\psi}_+, \qquad
\partial_x ( \sqrt{x} \tilde{\psi}_+ ) = i m \sqrt{x} \tilde{\psi}_-  \, . 
\label{d6}
\eeq
The second equation is readly  integrated 
\beq
\tilde{\psi}_+(x, t)   =  - \frac{ i m }{\sqrt{x} } \int_x^\infty dy \, \sqrt{y} \tilde{\psi}_- (y,t)  \, , 
\label{d7}
\eeq
and replacing it    into  the first equation in (\ref{d6}) gives 

\beq
i \partial_t \,  \tilde{\psi}_- (x,t)  = - i \sqrt{x} \partial_x  \left( \sqrt{x} \tilde{\psi}_- \right)  
-   i m^2  \sqrt{x}   \int_x^\infty    dy \, \sqrt{y}  \,   \tilde{\psi}_-    (y,t)   \, . 
\label{d8} 
\eeq
This is the Schroedinger equation with Hamiltonian (\ref{5b}) and  the relation   $m = \ell_p$
found in the previous  section. The non locality of the Hamiltonian
(\ref{5b}) is a consequence of the  special  coordinates $t,x$ where the component 
$\tilde{\psi}_+$ becomes non dynamical and depends non locally  on  the  component
$\tilde{\psi}_-$ that is identified with the wave function of the modified $xp$ model. 
Similarly, the boundary condition (\ref{h555}) can  be derived from the Eq.(\ref{d1b}) as follows. 
In Rindler coordinates the latter  equation reads
\barray 
e^{ - \phi}  \psi_-^\dagger (\ell_x, \phi)  \;  \delta \psi_-(\ell_x, \phi)   =   e^{ \phi}  \psi_+^\dagger (\ell_x, \phi)  \;  
\delta \psi_+(\ell_x, \phi) , \quad  \forall \phi \, , 
\label{d9}
\earray 
that is solved by  
\barray 
- i e^{i \vartheta}  \,   e^{ -  \phi/2}  \,  \psi_-(\ell_x, \phi)   =    e^{ \phi/2} \,   \psi_+(\ell_x, \phi), \quad  \forall \phi \, , 
\label{d100}
\earray
where $\vartheta \in [0, 2 \pi)$.  Using Eq.(\ref{d5}) this equation becomes 
\barray 
- i e^{i \vartheta}  \,   \tilde{\psi}_-(\ell_x, t)   =      \tilde{\psi}_+(\ell_x, t), \quad  \forall t, 
\label{d11}
\earray
that   together with Eq.(\ref{d7}) yields Eq.(\ref{h555}). 
This completes the derivation of the quantum Hamiltonian and boundary condition associated
to  $H = x( p + \ell_p^2/p)$. The eigenfunctions  and eigenvalue equation of this model were found  in Section V.  
However, we shall  rederive  them  in alternative way that will provide  new insights in the next section.

Let us start by  constructing the plane wave solutions of the Dirac equation  (\ref{d3}), 
\beq
\left( \begin{array}{c}
\psi_-  \\
\psi_+ \\
\end{array}
\right)  \propto 
\left( \begin{array}{c}
e^{i \pi/4}  e^{ \beta/2} \\
e^{-i \pi/4}  e^{ - \beta/2} \\
\end{array}
\right) e^{  i ( - p^0  x^0  + p^1  x^1) } \, , 
\label{d10}
\eeq
where $(p^0, p^1)$  is the energy-momentum vector parameterized in terms of the rapidity  variable 
 $\beta$
\barray
& &     (p^0)^2 - (p^1)^2 = m^2, \label{d11v} \\
p^0 & = &   i m \sinh \beta, \quad p^1 = i m \cosh  \beta, \qquad \beta \in (- \infty, \infty). 
\nonumber 
\earray  
In  Rindler coordinates  these   plane wave solutions decay  exponentially with the distance as corresponds to a localized
wave function 
\beq
e^{  i ( - p^0  x^0  + p^1  x^1) } = e^{ - m \rho  \cosh( \beta - \phi)} \rightarrow 0, \quad {\rm as} \quad \rho \rightarrow 	\infty \, . 
\label{d12}
\eeq
The general solution of the Dirac equation  is  given by the  linear  superposition
of plane waves (\ref{d10}). The superposition  that reproduces the  eigenfunctions of the modified $xp$ model is 
\barray 
\psi_\mp(\rho, \phi) & = & e^{ \pm i \pi/4}   \int_{- \infty}^\infty d \beta \, e^{-  i E \beta/2}  \,   
e^{  \pm \beta/2} \,  e^{ - m \rho  \cosh( \beta - \phi)} \,  
 \label{d13} \\
 & = & 2 e^{ \pm i \pi/4} e^{  ( \pm \frac{1}{2} -   \frac{i E}{2}  ) \phi} \,  K_{  \frac{1}{2} \mp    \frac{i E}{2} }( m \rho)  \, , 
 \nonumber 
\earray 
that replaced in Eq.(\ref{d100}) gives 
\beq
 e^{ i \vartheta} \, K_{ \frac{1}{2} -   \frac{ i E}{2}   } (m \ell_x)  -  K_{ \frac{1}{2} +  \frac{ i E}{2}} (m \ell_x) = 0 \, , 
\label{d14}
\eeq
which coincides with   the eigenvalue equation (\ref{h7}) with $m=\ell_p$.  Setting  
$m \ell_x = 2 \pi$ and $\vartheta = \pi$,   brings  Eq.(\ref{d14}) to the form 
\beq
\xi_{H} (t) \equiv 
 K_{ \frac{1}{2} +   \frac{ i t}{2}   } (2 \pi)  +  K_{ \frac{1}{2} -   \frac{ i t}{2}} (2 \pi) = 0. 
\label{d15}
\eeq

\vspace{0.25 cm}

{\bf Summary:}  

\vspace{0.25 cm}

\fbox{
\begin{minipage}{48em}

\checkmark The spectrum of a relativistic massive  fermion in the domain ${\cal U}$
agrees  with the average Riemann zeros.  

?  \;  Does this result  provide a hint on a physical realization  of the Riemann zeros. 

\end{minipage}
}

\section{$\xi$-functions: P\'olya's is massive and Riemann's is massless}

\begin{figure}[h!]
\begin{center}
\includegraphics[width=.4\linewidth]{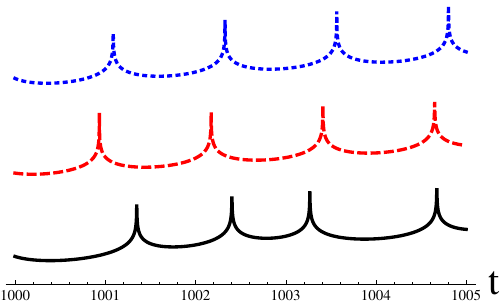}
\end{center}
\caption{From bottom to top: plot of $- \log |\xi(t)|$  (Riemann
zeros), $-  \log |\xi_H(t) |$ (eigenvalues of the Hamiltonian (\ref{5b}) with $\ell_x \ell_p = 2 \pi$) and 
$- \log  |\xi^*(t)|$ (P\'olya  zeros).  The cusp represents the zeros of the corresponding  functions.
  }
\label{xi}
\end{figure}

The function $\xi_H(t)$ appearing in  Eq.(\ref{d15}) reminds the {\em fake} $\xi$  function defined  by P\'olya in 1926 \cite{P26,H90}
\beq
\xi^*(t) = 4 \pi^2 \left( K_{ \frac{9}{4} + \frac{i t}{2}}( 2 \pi)  + K_{ \frac{9}{4} - \frac{i t}{2}}(2 \pi)  \right) \, . 
\label{po1}
\eeq 
This function  shares  several   properties with the Riemann  $\xi$ function
\beq
\xi(t) = \frac{1}{4} s (s-1) \Gamma \left( \frac{s}{2} \right) \pi^{- s/2} \zeta(s), \quad s= \frac{1}{2} + i t \, , 
\label{po2}
\eeq
namely, $\xi^*(t)$ is an  entire and  even function of $t$, its zeros lie on the real axis and  behave asymptotically like  the
average Riemann zeros,  as shown by the expansion obtained using Eq.(\ref{h8}) 
\beq
\xi^*(t)  \stackrel{t \rightarrow \infty}{\longrightarrow} 2^{3/4} \pi^{ - 7/4} t^{7/4} e^{ - \pi t/4} 
\cos \left( \frac{t}{2} \log \left( \frac{t}{ 2 \pi e} \right) 
+ \frac{ 7 \pi}{8} \right)   \, . 
\label{po3}
\eeq
The zeros of $\xi(t), \xi_H(t)$ and $\xi^*(t)$ are plotted in fig.\ref{xi}. 
The slight displacement  between the two top  curves  
is due to  the constant $7 \pi/8$ appearing in the argument of the cosine function 
in Eq.(\ref{po3}) as compared to that in Eq.(\ref{h9}). 

\begin{figure}[h!]
\begin{center}
\includegraphics[width=.4\linewidth]{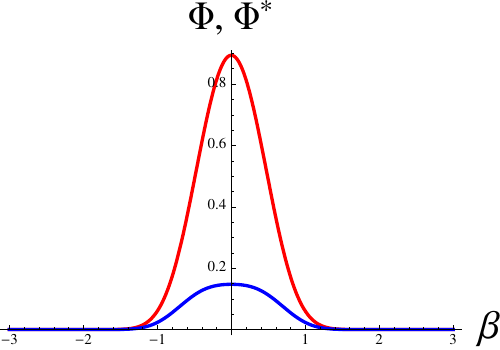}
\end{center}
\caption{Plot of $\Phi(\beta)$ (red on line), and $\Phi^*(\beta)$ (blue on line). Outside the region
$|\beta| <1$ the difference is very small. 
  }
\label{PolyaPhi}
\end{figure}

The similarity between $\xi_H(t)$ and $\xi^*(t)$,  and the relation between  $\xi^*(t)$ and 
$\xi(t)$ provides  a  hint on the field theory underlying the Riemann zeros.  To show this, we shall 
review how P\'olya arrived at  $\xi^*(t)$.  The starting point is the expression of
$\xi(t)$ as a Fourier transform \cite{T86}
\barray 
\xi(t) &  = &  4 \int_1^\infty dx \frac{ d [ x^{\frac{3}{2}} \psi'(x)]}{dx} \, x^{- \frac{ 1}{4}} \cos \left( \frac{t \log x}{2} \right) \, , 
 \label{po4} \\
\psi(x) &  = &  \sum_{n=1}^\infty e^{ - n^2  \pi x}, \quad \psi'(x) = \frac{d \psi(x)}{dx}   \, . 
\nonumber 
\earray 
In the variable  $x= e^{\beta}$  these  equations become,  
\barray 
\xi(t)  & =  &   \int_0^\infty d\beta  \;  \Phi(\beta) \cos \frac{ t \beta}{2}, \label{po5} \\
\Phi(\beta) &  = &  2 \pi e^{ 5 \beta/4} \sum_{n=1}^\infty \left( 2 \pi e^\beta n^2 - 3 \right) \, n^2 e^{ - \pi n^2 e^\beta}. 
\nonumber 
\earray 
The  function $\Phi(\beta)$  behaves asymptotically as 
\beq
\Phi(\beta) \rightarrow  4 \pi^2  e^{ 9 \beta/4}  e^{ - \pi e^\beta}, \qquad \beta \rightarrow \infty, 
\label{po6}
\eeq
which P\'olya replaced by the following expression   (see Fig. \ref{PolyaPhi}). 
\beq
\Phi^*(\beta)   =  4 \pi^2   \left( e^{ 9 \beta/4}  + e^{-  9 \beta/4} \right)   e^{ - \pi ( e^\beta + e^{- \beta})}.
\label{po7}
\eeq
The function $\xi^*(t)$ is defined as the Fourier transform of $\Phi^*(\beta)$, 
\beq
\xi^*(t) =  \int_0^\infty d\beta  \;  \Phi^*(\beta) \cos \frac{ t  \beta }{2} \, , 
\label{po8}
\eeq
which gives finally the Eq.(\ref{po1}).   The function (\ref{d15})
can also be written as the Fourier transform 
\beq
\xi_H(t) =  \int_0^\infty d\beta  \;  \Phi_H(\beta) \cos \frac{ t  \beta }{2}, 
\label{po9}
\eeq
with
\beq
\Phi_H(\beta) = ( e^{ \beta/2} + e^{ - \beta/2} ) e^{ - 2  \pi  \cosh \beta} 
\label{po10}
\eeq
Observe that the  term $e^{ - 2  \pi  \cosh \beta}$ appears  in $\Phi_H(\beta)$ and $\Phi^*(\beta)$. 
The origin of this term in the Dirac theory is the plane wave factor  (\ref{d12}) of a fermion with mass $m$
located  at the boundary $\rho=\ell_x$ with  $m \ell_x = 2 \pi$.  This observation suggests 
 that  the P\'olya $\xi$  function arises in the  relativistic  theory of a massive particle 
with scaling dimension  $9/4$, rather than  $1/2$, that  corresponds to a fermion
(this would  explain the different order  of  the corresponding Bessel functions).  
The approximation   $\Phi(\beta) \simeq  \Phi^*(\beta)$, that is 
$e^{ - \pi e^ {\beta}} \simeq  e^{ - 2 \pi \cosh \beta}$,  can   then  be understood as the 
 replacement of  a massless particle by a massive one.
Indeed, the energy-momentum of a massless right moving particle is given by $p^0 = p^1 =  \Lambda e^\beta$,
where $\Lambda$ is an energy scale.  The corresponding plane wave factor is   $e^{ - \pi e^ {\beta}}$,
with $\Lambda = \pi$. For large rapidities, $\beta \gg 1$, a massive particle behaves as a  massless one, i.e. 
$ e^{ - 2 \pi \cosh \beta} \simeq e^{ - \pi e^ {\beta}}$.  However,  for small  rapidities this is not the case.
These arguments suggest  that the field theory underlying the Riemann $\xi$ function, if it exists,
must  associated to a massless particle. 

\vspace{0.25 cm}

\vspace{0.25 cm}

{\bf Summary:} 

\vspace{0.25 cm}

\fbox{
\begin{minipage}{50em}

\checkmark The zeros of the Polya  $\xi^*$ function behave as the  spectrum of  a  relativistic massive particle in the  domain ${\cal U}$.

? \, Polya's construction of $\xi^*$ suggests that the Riemann's   $\xi$ function is  related to a massless particle.



\end{minipage}
}

\section{The massive Dirac Model  in Rindler coordinates}

Let us formulate  the Dirac theory in Rindler coordinates. 
Under a Lorentz  transformation with boost parameter $\lambda$,  the 
light cone coordinates  $x^\pm$ and the Dirac spinors $\psi_\pm$    transform as
\beq
x^\pm \rightarrow e^{\mp \lambda} \, x^\pm, \qquad \psi_\pm \rightarrow e^{\pm  \lambda/2} \, \psi_\pm,
\label{R351}
\eeq
and  the Rindler coordinates  (\ref{ri1})  as 
\beq
\phi \rightarrow \phi - \lambda, \qquad \rho \rightarrow \rho.
\label{R352}
\eeq
Hence the new spinor fields  $\chi_\pm$ defined as 
\beq
\chi_\pm = e^{ \pm \phi/2} \, \psi_\pm,
\label{R40}
\eeq
remain   invariant under (\ref{R352}).  The  Rindler wedge ${\cal R}_+$,  and its domain ${\cal U}$, 
are also invariant under Lorentz  transformations. 
The Dirac action (\ref{d0}) written  in terms of the spinors $\chi_\pm$ reads 
\barray 
S   & = &   \frac{i}{2}
\int_{- \infty}^\infty d \phi \, \int_{\ell_x}^{\infty} d \,   \rho \left[ \chi^\dagger_- (  \partial_\phi + \rho \partial_\rho  + \frac{1}{2} )  \chi_-    
+  \chi^\dagger_+ ( \partial_\phi -  \rho \partial_\rho  -   \frac{1}{2} )  \chi_+  +   i m  \rho \left(  \chi^\dagger_-  \chi_+   +  \chi^\dagger_+ \chi_-   \right) 
\right]  \, , 
\label{Sd1}
\earray 
while the Dirac equation (\ref{d1}) and the boundary condition (\ref{d100}) become
\beq
 ( \partial_\phi  \pm  \rho  \partial_\rho \pm \frac{1}{2} ) \chi_\mp  +i m  \rho  \chi_\pm = 0, 
\label{R41}
\eeq
and
\barray 
- i e^{i \vartheta}  \,     \chi_-    =       \chi_+  \quad {\rm at} \; \;  \rho = \ell_x  \, . 
\label{R50}
\earray
 The infinitesimal generator of translations of the Rindler time $\phi$, 
acting on the spinor wave functions,  is the Rindler Hamiltonian $H_R$, that 
can be read off from (\ref{R41})
\beq
i \partial_\phi \chi = H_R \, \chi, \qquad \chi  = \left( 
\begin{array}{c}
\chi_- \\
\chi_+ \\
\end{array}
\right), 
\label{R42}
\eeq

\beq
H_R = 
\left( \begin{array}{cc}
- i ( \rho \,   \partial_\rho + \frac{1}{2} )  & m \rho \\ 
m \rho &  i ( \rho \,   \partial_\rho + \frac{1}{2} ) \\
\end{array}
\right) = \sqrt{\rho} \, \hat{p}_\rho \, \sqrt{\rho} \, \sigma^z + m \rho \,  \sigma^x , 
\label{R43}
\eeq
where  $\hat{p}_\rho = - i \partial / \partial \rho$, is the momentum operator conjugate 
to the radial coordinate $\rho$. Notice  that  the operator 

\beq
H_{\rho p_\rho} = - i ( \rho \,   \partial_\rho + \frac{1}{2} ) = \frac{1}{2} ( \rho \, \hat{p}_\rho + \hat{p}_\rho \rho) =
\sqrt{\rho} \, \hat{p}_\rho \, \sqrt{\rho}, 
\label{R43b}
\eeq
coincides with Eq.(\ref{qu1})  with the identification  $x = \rho$  (in units $\hbar =1$). 
The  eigenfunctions  of (\ref{R43b}) are 

\beq
H_{\rho p_\rho}\, \psi_E = E \, \psi_E, \qquad \psi_E = \frac{1 }{ \sqrt{2 \pi}} \rho^{- 1/2 + i E}, 
\label{Rxpb}
\eeq
with real eigenvalue $E$  for    $\rho >0$ (recall Eq.(\ref{qu6})). 
Thus $H_R$ consists of two copies of $xp$,  with different signs corresponding to 
opposite fermion  chiralities that are coupled by the  mass term $m \rho \sigma^x$. 

The scalar product of two wave functions, in the domain ${\cal U}$,  can  be defined as 
\beq
\langle \chi_1 | \chi_2 \rangle = \int_{\ell_x}^\infty  d \rho 
  \left(  \chi^*_{1,-} \chi_{2,-}   +  \chi^*_{1,+} \chi_{2,+}   \right). 
  \label{R431}
  \eeq
The  Hamiltonian $H_R$ is hermitean with this scalar product acting on wave functions
that  satisfy Eq.(\ref{R50}) and vanish sufficiently fast at infinity, i.e. 
$\lim_{\rho \rightarrow \infty} \rho^{1/2} \chi_\pm(\rho, \phi) =0$. 
The eigenvalues and eigenvectors of the  Hamiltonian (\ref{R43}), are given by the solutions of the 
Schroedinger equation 
\beq
H_R \,  \chi = E_R  \, \chi, \qquad 
 \chi_\pm(\rho, \phi)  =  e^{ - i E_R  \phi  \, \mp i \pi/4}   K_{\frac{ 1}{2}  \pm i E_R}(m \rho),, \qquad  \rho \geq \ell_x, 
\label{R44}
\eeq
which coincide with Eq.(\ref{d13}) with the identification

\beq
E_R = \frac{E}{2} \, . 
\label{R444}
\eeq
The factor of $1/2$ comes from the relation $e^{ 2 t} =  x^0+ x^1 = \rho e^\phi$ (see  Eq.(\ref{11b})), 
that implies $e^{ - i E_R \phi} \propto e^{ - i E t}$.  The Rindler eigenenergies are obtained
replacing $E$ by $2 E_R$ in Eq.(\ref{d14}). 

\vspace{0.25 cm}

{\bf Comments: }

\vspace{0.25 cm}

\begin{itemize}

\item    The Dirac Hamiltonian   associated to the metric (\ref{ri5}) is 
\beq
H  = \left( 
\begin{array}{cc}
h   & m \rho \Lambda  \\
m \rho \Lambda  & -h    \\
\end{array}
\right), \qquad h = - i   \left( \rho \partial_\rho + \frac{1}{2} + \frac{1}{2} \rho \partial_\rho (\log \Lambda) \right) , \qquad 
\Lambda = \frac{ \rho}{\sqrt{\rho^2 - \ell_x^2}} \, . 
\label{R555}
\eeq
%
In the limit $\rho \gg \ell_x$ this Hamiltonian converges towards  (\ref{R43}). 

\item Gupta,  Harikumar and de Queiroz
 proposed  the Hamiltonian $( x {\slashed p} + {\slashed p} x)/2$ 
as a Dirac variant of the $xp$ Hamiltonian \cite{G12}. The Hamiltonian is defined on a semi-infinite cylinder
and becomes effectively one dimensional by considering the winding modes on the compact  dimension. 
The eigenfunctions are given by  Whittaker functions and the  spectrum satisfies an equation
similar to Eq.(\ref{L20}) in  the Landau theory. In the limit where a regularization parameter goes
to zero one obtains a continuum spectrum with a correction term related to the Riemann-von Mangoldt
formula.

\item Bender, Brody and M\"{u}ller  proposed recently a generalization of the $xp$ operator \cite{BB16}
\beq
H = \frac{ \bf 1}{ {\bf 1}  - e^{ - i \hat{p} }}  ( x \hat{p} + \hat{p} x )   ( {\bf 1}   - e^{ - i \hat{p}} ) \, , 
\label{bb1}
\eeq
with the property that its eigenvalues $E_n$ give the Riemann zeros as $z_n = \frac{1}{2} ( 1 - i E_n)$. 
This interesting result follows from the fact the eigenfunctions of (\ref{bb1}) are given in terms of the  Hurwitz
zeta function as  $\psi_z(x) = \zeta(z, x+1)$ and   imposing the boundary condition 
\beq
\psi_{z_n}(0) =0  \rightarrow \zeta(z_n, 1) = \zeta(z_n) = 0 \, . 
\label{bb2}
\eeq
Unfortunately  the operator (\ref{bb1}) is not self-adjoint, so that the reality of its eigenvalues
is not guaranteed. However,  the authors of \cite{BB16} found that $i H$ has a $PT$ 
symmetry which, if it is maximally broken,  would imply the reality of the eigenvalues. 
This property though remains to be proved. Further details can be found in references
\cite{B17,BB17}.


%

\end{itemize}

\vspace{0.25 cm}

{\bf Summary:} 

\vspace{0.25 cm}

\fbox{
\begin{minipage}{50em}

\checkmark The massless Dirac Hamiltonian in Rindler spacetime is the direct sum of $xp$  and $- xp$. 

\checkmark The mass term couples  the left and right modes of the fermions.


\end{minipage}
}

\vspace{1cm}

\section{The massless Dirac equation with  delta function potentials}

\begin{figure}[t]
\begin{center}
{\includegraphics[height= 6.0 cm]{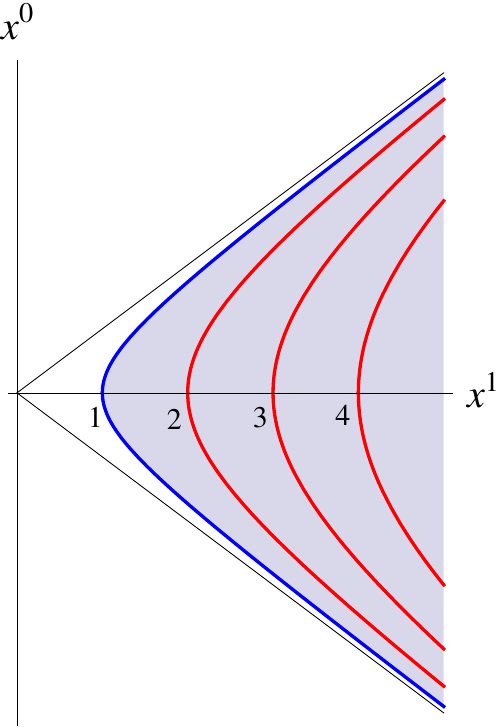}
\hspace{1.5cm}
\includegraphics[height= 5.5 cm]{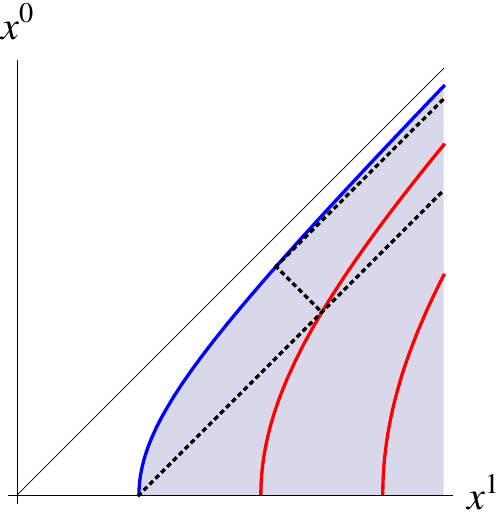}}
\end{center}
\caption{Left: worldlines of the mirrors with accelerations $a_n=  1/\ell_n = 1/n  \; (n=1,2, \dots )$. 
Right:  A massless fermion  (dotted line)  at the point $(x^0, x^1)= (0,1)$ moves to the right until 
it hits a moving  mirror where it can be reflected or transmitted. 
} 
\label{moving-mirror}
\end{figure} 

From analogies between the Polya $\xi^*$ function, the  Riemann $\xi$ function  and 
the $\xi_H$ function of the  massive Dirac model,  we conjectured
 in section VIII  the existence of a  massless field theory underlying $\xi$. 
 At first look this  idea does not look  correct   because
the  Hamiltonian obtained by setting $m=0$ in Eq.(\ref{R43}), 
is equivalent to two copies of the quantum $xp$ model  which
has a continuum spectrum. In fact, the  mass  term in that Hamiltonian is the mechanism responsible for 
obtaining a discrete spectrum.

To resolve this puzzle we shall replace the {\em bulk} mass term in the Dirac action
(\ref{Sd1}) by a sum of ultra-local interactions  placed  at fixed values  $\ell_n$  of the radial coordinate $\rho$ \cite{S14}. 
These interactions can arise from  moving mirrors, or beam splitters, that move with a  uniform acceleration $1/\ell_n$
(see Fig. \ref{moving-mirror}).  The fermion moves freely, until it hits one of the mirrors   and it is 
 reflected or transmitted.  The moving mirrors  are realized  mathematically by 
 delta functions added to  the  massless Dirac action  that couple the left  and right components of the fermion on both sides
of the mirror.  These  delta functions provide the matching  conditions for  the wave functions  and 
 can be  parameterized by a complex number  $\varrho_n$ with  $n=2, \dots, 	\infty$. 
The scattering of the fermion at each mirror preserves  unitarity that is equivalent to the self-adjointness of the Hamiltonian. 

\begin{figure}[h!]
\begin{center}
\includegraphics[height= 2.2 cm]{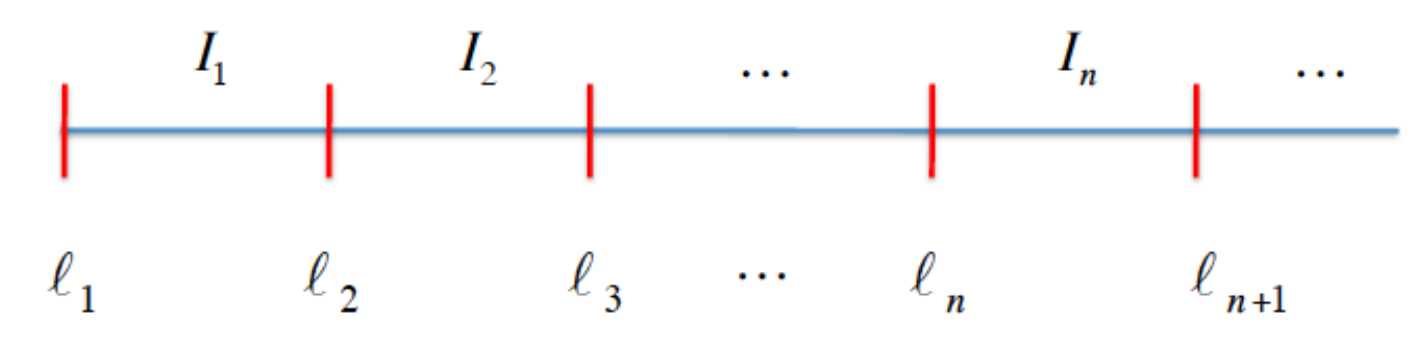}
\end{center}
\caption{Intervals  $I_n$ defined in Eq.(\ref{m2}) 
} 
\label{recta}
\end{figure}

The model  is formulated in the  spacetime  ${\cal U}$ defined in Eq.(\ref{uspace}). 
 We divide  ${\cal U}$ into an infinite number of domains separated by hyperbolas with constant
 values of $\rho= \ell_n$, as follows. First we define the intervals  (see Fig.\ref{recta}) 
\barray
  {I}_n =\{ \rho\, | \; \ell_n < \rho < \ell_{n+1} \}, \quad n= 1,2, \dots, \infty  \, , 
\label{m2}
\earray 
where using the scale invariance of the model we set  $\ell_1 =1$
($\ell_1$ plays the role of $\ell_x$ in previous sections).

The partition  of  ${\cal U}$ is given by 
\barray
{\cal U} & \rightarrow &  \tilde{{\cal U}}   =  \cup_{n=1}^\infty  \, {\cal U}_n, 
 \qquad {\cal U}_n = {\cal I}_n \times \Rmath \, , 
\label{m4}
\earray 
where the   factor $\Rmath$  denotes  the range  of the Rindler time $\phi$. See Fig. \ref{moving-mirror} for an  example with  $\ell_n =n$. 
The  wave function of the model is  the two  component Dirac spinor  (see Eq.(\ref{R42}))  
\beq
 \chi(\rho)   = \left( 
\begin{array}{c}
\chi_-(\rho)  \\
\chi_+(\rho)  \\
\end{array}
\right), \qquad  \rho \in {\cal I} = \cup_{n=1}^\infty I_n,  
\label{m5}
\eeq
and  the  scalar product  is  given by   (recall Eq.(\ref{R431})) 
\barray 
\langle \chi | \chi \rangle  &  =  &     \sum_{n=1}^\infty  \int_{\ell_n}^{\ell_{n+1}}  d \rho \,   \chi^\dagger (\rho) \cdot  \chi(\rho) . 
\label{m6}
\earray 
%
The complex Hilbert space  is ${\cal H}_{} = L^2({\cal I}, \Cmath) \oplus L^2({\cal I}, \Cmath)$ and 
the Hamiltonian  is obtained setting  $m=0$ in Eq.(\ref{R43}) 
\beq
H = 
\left( \begin{array}{cc}
- i ( \rho \,   \partial_\rho + \frac{1}{2} )  & 0  \\ 
0  &  i ( \rho \,   \partial_\rho + \frac{1}{2} ) \\
\end{array}
\right) , \quad \rho \notin {\cal I}  \, . 
\label{m8}  
\eeq
$H$ is a self-adjoint operator acting on the subspace  ${\cal H}_\vartheta \subset   {\cal H}$  of 
wave functions that satisfy the boundary conditions  \cite{S14}  
(see \cite{AIM05} for the relation between self-adjointness of operators and boundary conditions)

\begin{figure}[h]
\begin{center}
\includegraphics[height= 3.2 cm]{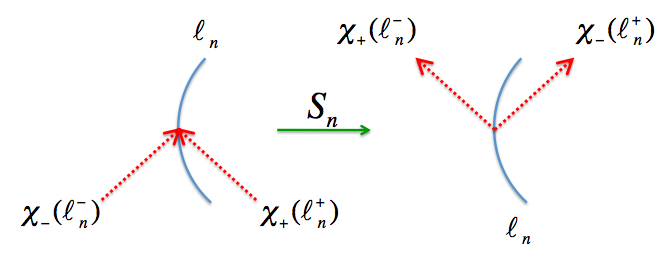}
\hspace{1.5cm}
\includegraphics[height= 3.2 cm]{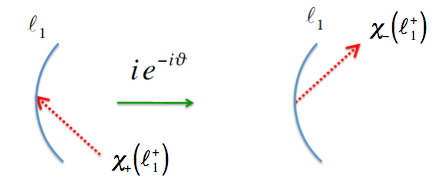}
\end{center}
\caption{Left: scattering process taking place at the mirror located at $\rho = \ell_n$ for $n>1$ (Eq.(\ref{m13})). Right:
 reflexion at the  perfect mirror located at  $\rho = \ell_1$ (Eq.(\ref{m9})). 
} 
\label{S-matrix}
\end{figure}

\barray 
\chi  & \in &  {\cal H}_{\vartheta}: \quad \chi(\ell_n^-)  =  L(\varrho_n)  \, \chi(\ell_n^+),  \quad (n  \geq 2), \quad 
- i e^{i \vartheta}   \,  \chi_-(\ell_1^+)    =   \chi_+(\ell_1^+),  
\label{m9}   
\earray
where 
\beq
\chi(\ell_n^\pm) = \lim_{\varepsilon \rightarrow  0^+} \chi(\ell_n \pm \varepsilon) , 
\label{m10}
\eeq
and 
\barray
 \vartheta  \in [0, 2  \pi), \qquad 
{L}(\varrho)   & = &   \frac{1}{1 - |\varrho|^2}  
\left( \begin{array}{cc}
1 + |\varrho|^2    &   2 i  \varrho\,     \\
- 2 i   \varrho^*   & 1 + |\varrho|^2    \\
\end{array}
\right), \quad \varrho   \in \Cmath, \quad |\varrho|   \neq 1.    \label{5}  
\label{m11} 
\earray 
This means  that  $H$ satisfies 
\barray 
\langle \chi_1 | H    \chi_2 \rangle  =  \langle H   \chi_1 |\chi_2 \rangle,  \qquad \chi_{1,2} \in {\cal H}_{\vartheta}. 
\label{m11b}
\earray 
This condition guarantees that the norm (\ref{m6}) of the state  is conserved by the time evolution generated by the Hamiltonian. 
The subspace ${\cal H}_\vartheta$ also depends on  $\ell_n$ and $\varrho_n$  but we shall not write
this dependence explicitly. Similarly, we shall also denote  the Hamiltonian  as $H_\vartheta$. 
The matching conditions  (\ref{m9}) describe  a scattering process where two  incoming waves   $\chi^{\rm in}_n$ 
 collide at the  $n^{\rm th}$-mirror and become  two   outgoing waves  $\chi^{\rm out}_n$ given by  (see Fig. \ref{S-matrix}) 
\beq
\chi^{\rm in}_n   = \left( 
\begin{array}{c}
\chi_-(\ell_n^-)  \\
\chi_+(\ell_n^+)  \\
\end{array}
\right) , \quad 
\quad \chi^{\rm out}_n   = \left( 
\begin{array}{c}
\chi_-(\ell_n^+)  \\
\chi_+(\ell_n^-)  \\
\end{array}
\right),  \, \quad  n >  1 \, . 
\label{m12} 
\eeq
At the mirror  $n=1$, the components $\chi_\pm(\ell_1^-)$ of these vectors  are null, i.e.
there is no propagation at the left of the boundary.  
The scattering process  is described by  the matrix $S_n$ 
\beq 
\chi^{\rm out}_n = S_n \, \chi^{\rm in}_n , \qquad S_n = \frac{1}{1+ |\varrho_n|^2} 
\left( \begin{array}{cc}
1 - | \varrho_n|^2 & - 2 i \varrho_n  \\
- 2 i  \varrho_n^* & 1 -| \varrho_n|^2 \\
\end{array}
\right) , \, \quad  n >  1 \, ,  
\label{m13}
\eeq
that  is unitary,
\beq
S_n \, S^\dagger_n = {\bf 1} \, . 
\label{m14}
\eeq
Notice that the  boundary condition at $\rho = \ell_1$, is  also described by  Eq.(\ref{m13}) with a parameter $\varrho_1$ 
\beq
\varrho_1 = - e^{ - i \vartheta} \, , 
\label{m13b}
\eeq
that is a pure phase for the Hamiltonian $H_\vartheta$ to be self-adjoint. 
The matrix  $L(\varrho)$ satisfies 
\beq
L(1/\varrho^*) = - L(\varrho).
\label{m15}
\eeq
Hence,  replacing $\varrho_n$ by $1/\varrho_n^*$ gives a unitary equivalent  model
because the sign change at $\rho = \ell_n$, given
in Eq.(\ref{m15}), can be compensated by changing the sign 
of the wave function in the remaining   intervals.  
Hence, without losing  generality,  we shall   impose
the condition $|\varrho_n| < 1, \; \forall n >  1$.  

%


The eigenfunctions of  the Hamiltonian (\ref{m8})   are the customary  functions  (see Eq.(\ref{qu6}))
\beq
H  \, \chi = E \, \chi  
\longrightarrow  \chi_\mp \propto \rho^{ - 1/2  \pm i E}  \, . 
\label{m16}
\eeq 
From now one, we shall assume that $E$ is  a real number  which   is guaranteed by the self-adjointness of the Hamiltonian $H$. 
In  the $n^{\rm th}$ interval  we take
\beq
\chi_{\mp, n}(\rho) = e^{ \pm i \pi/4}  \frac{A_{\mp, n}}{ \rho^{  1/2 \mp  i E}},    \qquad  \ell_n < \rho < \ell_{n+1},  
\label{m17}
\eeq
where $A_{\mp, n}$ are constants that, in general,  will depend on $E$.  The phases $e^{\pm i \pi/4}$ have been introduced 
by analogy with those appearing in Eq.(\ref{R44}). The boundary values of $\chi$ at  $\rho= \ell_n^\pm  \; (n \geq 1)$ are  
(see Eq.(\ref{m10}))
\barray 
   \chi_\mp(\ell^+_n)   =    \chi_{\mp, n}(\ell_n) = 
e^{ \pm \frac{ i \pi}{4}}    \frac{A_{\mp, n}}{ \ell_n^{1/2  \mp  i E } }, \qquad 
\chi_\mp (\ell^-_n)   = 
\chi_{\mp, n-1}(\ell_n) =  e^{ \pm \frac{ i \pi}{4}}   \frac{A_{\mp, n-1}}{ \ell_n^{ 1/2  \mp  i E } } . 
\label{m18}  
\earray 
Let us define   the vectors 
\beq
| {\bf 	A}_n   \rangle = 
\left( \begin{array}{c}
A_{-, n} \\
A_{+, n}  \\
\end{array}
\right),   \qquad n \geq 1. 
\label{m19}
\eeq
The boundary conditions (\ref{m9}) together with Eqs.(\ref{m18})  imply 
\barray 
 | {\bf 	A}_{n-1}   \rangle &  = &  T_n \, | {\bf 	A}_n   \rangle \quad  \; (n\geq 2), \qquad  | {\bf 	A}_1   \rangle = 
 | {\bf 	A}_1  (\vartheta)   \rangle = 
\left( \begin{array}{c}
1 \\
e^{i \vartheta}  \\
\end{array}
\right)   \, , 
\label{m20}
\earray 
where   the transfer matrix $T_n$ is given by 
\barray
{T}_{n}  & = &     
 \frac{1}{1 - |\varrho_n|^2}  
\left( \begin{array}{cc}
1 + |\varrho_n|^2    &   2  \varrho_n\,  \ell_n^{ - 2 i E}   \\
2  \varrho_n^* \, \ell_n^{ 2 i E}  & 1 + |\varrho_n|^2    \\
\end{array}
\right) \quad  (n \geq 2) . 
 \label{m21}
\earray 
The norm of the eigenstate can be computed using equations  (\ref{m6})  and  (\ref{m17})
\barray 
 || \chi ||^2  = 
\sum_{n=1}^\infty \log \frac{\ell_{n+1}}{ \ell_n}  \, \langle {\bf A}_n | {\bf A}_n \rangle,  \qquad  \langle {\bf A}_n | {\bf A}_n \rangle =     |A_{-,n} |^2 + |A_{+, n}|^2.
\label{m22}  
\earray 
The $\log$ term comes from the integral  of the norm of the wave function in the $n^{\rm th}$ interval, 
$\int_{\ell_n}^{\ell_{n+1}}  d \rho/\rho$ (we used that $E$ is real).  
If $\varrho_n =0$ then  $T_n = {\bf 1}$ which  implies that   $| {\bf A}_{n-1}  \rangle =  \, | {\bf A}_n \rangle$. If this happens
for all $n$,  then $ | {\bf A}_n \rangle =  | {\bf A}_1 \rangle$, in which case the  norm of these states  diverges, but they can be normalized
using Dirac delta functions, so they correspond to scattering states. 
 In the general case, iterating   Eq.(\ref{m20})  yields   $| {\bf A}_{n} \rangle$ 
in terms of  $| {\bf A}_1 (\vartheta)    \rangle$  
\barray 
| {\bf A}_{n}  \rangle  &   = &  T_n^{-1} T_{n-1}^{-1} \cdots T_2^{-1} | {\bf A}_{1} (\vartheta)  \rangle, \quad n \geq 2 \, . 
\label{m23}
\earray 
 For special values  of $\ell_n$ and $\varrho_n$ one can  find
the  exact expression of these amplitudes. An example is $\ell_n = e^{ n/2}, \varrho_n = {\rm cte}$  \cite{S14}. 
In order to make contact with the Riemann zeros, we shall consider a limit  where the reflection coefficients vanish asymptotically. 


\vspace{0.25 cm}

{\bf Summary:} 

\vspace{0.25 cm}

\fbox{
\begin{minipage}{50em}

\checkmark The massless Dirac Hamiltonian with delta function potential is solvable 
by transfer matrix methods. 

\checkmark The model is completely characterized by the set of parameters
$\{ \ell_n, \varrho_n \}_{n=2}^\infty$ and $\vartheta$. 



\end{minipage}
}

\section{Heuristic approach to  the spectrum}

\begin{figure}[t]
\begin{center}
{\includegraphics[height= 2.5 cm]{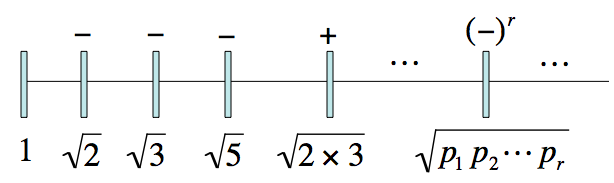}
}
\end{center}
\caption{Localization of the mirrors corresponding to the choice (\ref{m27}), together with the 
values of $\mu(n)$. 
} 
\label{A-mirror}
\end{figure}

\begin{figure}[t]
\begin{center}
{\includegraphics[height= 7. cm]{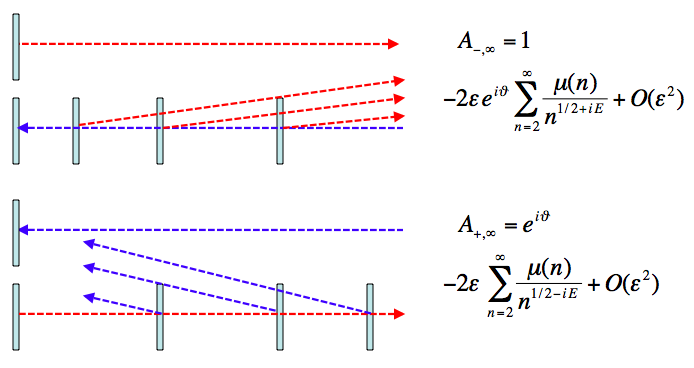}
}
\end{center}
\caption{Depiction of the  amplitudes $A_{\pm, \infty}$ as  the superposition of a principal wave with the
waves resulting from the  scattering with all the mirrors along its trajectory (see  Eq.(\ref{m26})). The terms of higher order in $\varepsilon$
correspond to more than one scattering.  
} 
\label{Ainfinity}
\end{figure}

Let us replace  $\varrho_n$ by $\varepsilon \varrho_n$, and consider the limit $\varepsilon \rightarrow 0$
of  the transfer  matrix  (\ref{m21}) 
\barray 
T_n & \simeq  & {\bf 1} + \varepsilon \,   \tau_n + O(\varepsilon^2), \qquad 
{\tau}_{n}  =   
\left( \begin{array}{cc}
0  &    2  \varrho_n\,   \ell_n^{ - 2 i E}   \\
2  \varrho_n^* \, \ell_n^{ 2 i E}  & 0   \\
\end{array}
\right) \qquad (n \geq 2)  . 
 \label{m24}
\earray 
 Plugging this equation  into Eq.(\ref{m23})  yields  
\barray 
| {\bf A}_{n}  \rangle   \simeq  \left( {\bf 1 } - \varepsilon \sum_{m=2}^n   \tau_m   \right)  | {\bf A}_{1} (\vartheta)  \rangle  + O(\varepsilon^2)
, \quad n \geq 2, 
\label{m25}
\earray 
and  in  components 
\barray 
A_{-, n} & \simeq & 1 - 2  \varepsilon \,   e^{ i \vartheta}  \sum_{m=2}^n \varrho_m  \, \ell_m^{ - 2 i E} + O(\varepsilon^2) , 
\qquad 
A_{+, n}  \simeq  e^{ i \vartheta}  -  2 \varepsilon \sum_{m=2}^n \varrho_m^*   \, \ell_m^{ 2 i E} + O(\varepsilon^2)  . 
\label{m26}  
\earray 
For a normalizable state,   the amplitudes   $A_{\pm, n}$ have to    vanish 
 as $n \rightarrow \infty$. In the next section we shall study in detail  the normalizability of the state. 
 We shall make the following choice  of   lengths and reflection coefficients    \cite{S14}
\beq
\ell_n = n^{1/2} , \qquad \varrho_n = \frac{\mu(n)}{n^{1/2}}  \, , \qquad n > 1, 
\label{m27}
\eeq
where $\mu(n)$ is the Mo\"ebius function that  is equal to $(-1)^r$,  
with  $r$ the number of distinct primes factors of a square free integer $n$,  and $\mu(n) =0$, if 
$n$ is divisible by  the square of a prime number \cite{D}. 
See figs. \ref{A-mirror} and \ref{Ainfinity} 
for a  graphical representation of Eqs.(\ref{m27}) and (\ref{m26}). 
The Moebius function has been used in the past to provide physical models of prime numbers,
most notably in the ideal gas of primons with fermionic statistics \cite{J90}-\cite{S90} 
and a potential whose semiclassical spectrum are the primes  \cite{M97,SH11}.

Another motivation of the choice (\ref{m27}) is the following \cite{S14}. 
Consider a fermion that leaves the boundary  at $\rho= \ell_1$, moves rightwards  until it hits the mirror at 
 $\rho= \ell_n$ where it gets reflected and returns  to the boundary. The time lapse for the entire trajectory is 
 given by 
 \beq 
 \tau_n = 2 \log (\ell_n/\ell_1) \, 
 \label{ti}
 \eeq
 where we used  the Rindler metric  Eq.(\ref{r1c}).  
If the mirror is associated to the prime $p$, that is $\ell_p = \sqrt{p}$,  the time will be  given by $\tau_p = \log p$.  
This result  reminds  the  Berry conjecture that postulates  the  existence of a classical chaotic Hamiltonian whose primitive periodic orbits
are labelled by the primes $p$, with periods $\log p$, and whose quantization will give the Riemann zeros as energy
levels \cite{B86}. A classical Hamiltonian with this property has not  been found, but the array of  mirrors  presented
above, displays some of its properties. In particular, the  trajectory  between  the boundary and the mirror at $\ell_p$, with $p$ a prime
number,   behaves as a  primitive orbit with a period $\log p$. Moreover, the trajectories and periods of these orbits  are independent of
the  energy of  the fermion because it moves at the speed of light.

Let us work out the consequences (\ref{m27}). 
The condition  for a normalizable  eigenstate, that is $\lim_{n \rightarrow \infty} A_{\pm, n} = 0$, is 
\beq
1\simeq  2 \varepsilon \, e^{i \vartheta}   \sum_{n=1}^\infty \frac{ \mu(n)}{ n^{ \frac{1}{2}  + i E} }  =  \frac{ 2 \varepsilon \, e^{i \vartheta} }{ \zeta( \frac{1}{2} + i E (\varepsilon))}, 
\label{m28}
\eeq
where we have included  the term $n=1$ in the series because it does not modify its value  when  $\varepsilon \rightarrow 0$. 
We have  employed the formula $\sum_{n =1}^\infty \mu(n)/n^s =  1/\zeta(s)$ for a value of  $s$  where the series  may not
converge.  In the next section we shall compute the value of the finite sum that determines  the norm of the state.  
$E_n(\varepsilon)$ denotes   a solution such that 
$\lim_{\varepsilon \rightarrow 0} E_n(\varepsilon) = E_n$, where 
$\frac{1}{2} + i E_n$ is a zero of the zeta function.  All known zeros of $\zeta(s)$ on the critical
line are simple, but we shall also consider the case where $\frac{1}{2} + i E_n$  might be a zero of order $r \geq 1$, 
that is $\zeta^{(r)}(s) \neq 0$. The Taylor expansion of $\zeta(\frac{1}{2} + i E(\varepsilon))$  around $\frac{1}{2} + i E_n$,
in eq.(\ref{m28}) yields
\beq
1\simeq   \frac{ 2 \varepsilon  \,  r!  \, e^{i \vartheta} }{ i^r (E_n(\varepsilon) - E_n) ^r  \zeta^{(r)}( \frac{1}{2} + i E_n)} \, . 
\label{m29}
\eeq
Hence $E_n(\varepsilon) - E_n$ is of order $\varepsilon^{1/r}$, as  $\varepsilon \rightarrow 0$
and 
\beq
 \frac{ \zeta^{(r)}( \frac{1}{2} + i E_n)}{ \zeta^{(r)}( \frac{1}{2} - i E_n)} = (-1)^r e^{ 2 i \vartheta} \, . 
 \label{m28f}
\eeq
On the other hand, from  Eq.(\ref{cl5c})  one finds  
 \beq
 i^r \, 
  \zeta^{(r)} ( \frac{1}{2} +  i E_n) =  e^{ -     i \theta(E_n)}  Z^{(r)}(E_n) \, ,
 \label{m29}
 \eeq
that plugged into (\ref{m28f}) yields
\beq
e^{ 2 i ( \vartheta+  \theta(E_n))} =  1,  \qquad \forall r 
\label{m29b}
\eeq

\begin{figure}[h]
\begin{center}
{\includegraphics[height= 4. cm]{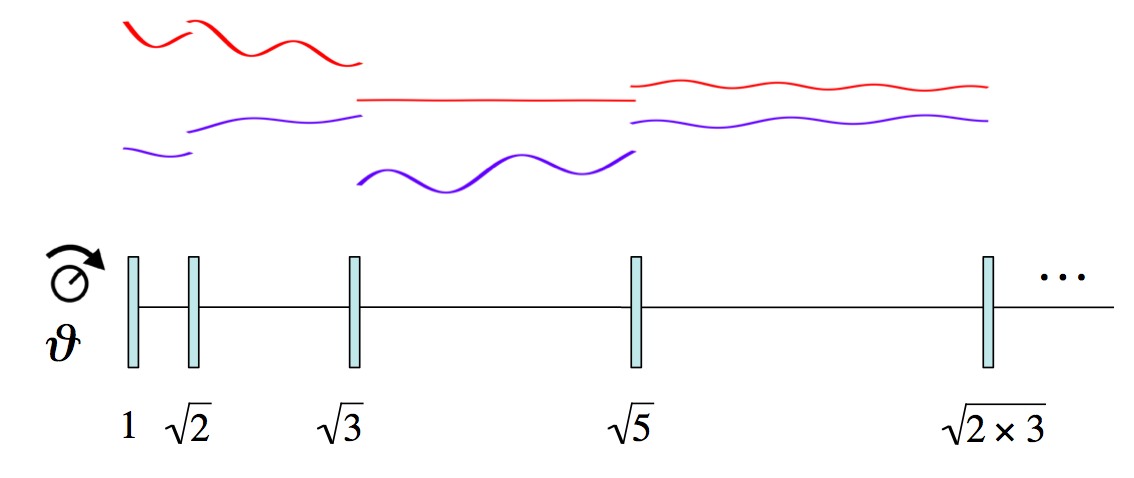}
}
\end{center}
\caption{Schematic representation of the array of mirrors that give rise to a spectral realization of the Riemann zeros. 
The red and blue  lines  represent the left and right wave functions $\chi_{\pm, n}(\rho)$. The wave functions  
are discontinuous at the moving mirrors located at the positions $\ell_n = \sqrt{n}$ with $n$ a square free integer. 
The knob on the  left represents the scattering phase at the perfect mirror that 
 is set to minus the  phase of the zeta function at the {\em zero} $E_n$, namely $\vartheta = - \theta(E_n) \; {\rm mod} \, \pi$. 
} 
\label{tune-zero}
\end{figure}

We can collect these results in the equation 
\beq
{\rm If} \;  \; \zeta ( \frac{1}{2} \pm i E_n ) = 0 \; \;  \:  {\rm and} \;  \;  	\; 
e^{ 2 i ( \vartheta+  \theta(E_n))} =  1 \Longleftrightarrow H_\vartheta \, \chi_{E_n}  = E_n \,  \chi_{E_n}  \, . 
\label{su}
\eeq
Observe that $\vartheta$ is fixed  mod $\pi$. In the next section we shall fix this ambiguity. 
This equation  is  heuristic. It has been derived  by i)  solving the eigenvalue equation in the
limit $\varepsilon \rightarrow 0$, ii)  imposing the vanishing of the eigenfunction at infinity and iii) 
using  the Dirichlet series of $1/\zeta(s)$ in a region where it may not converge. 
In the next section we shall derive Eq.(\ref{su}) without making the previous assumptions
(see  Eq.(\ref{r64})). 
Let us  notice that this spectral realization of the {\em zeros} 
requires  the fine  tuning of the parameter of $\vartheta$  in terms
of the phase of the zeta function, $\theta(E_n)$ (see fig. \ref{tune-zero}).  
This realization  is different from the P\'olya-Hilbert conjecture  of a single Hamiltonian encompassing
all the Riemann zeros at once. This    Hamiltonian would exist 
if $\theta(E_n) = \theta_0, \forall n$,   but this is certainly  not the case.

\vspace{0.55 cm}

{\bf Summary:} 

\vspace{0.25 cm}

\fbox{
\begin{minipage}{50em}

\checkmark A Riemann zero,  on the critical line,  becomes  an  eigenvalue
of the Hamiltonian $H_\vartheta$ by tuning the phase $\vartheta$ according to the 
phase of the zeta function. 

\vspace{0.2 cm}

\xmark \,   The previous result is   obtained in the  limit $\varepsilon \rightarrow 0$   and is  heuristic.

\end{minipage}
}

\vspace{0.2 cm}

\section{The Riemann zeros as spectrum  and The Riemann hypothesis}

In this section we provide  rigorous arguments  that support  the heuristic results obtained
previously. 
Let us first  review   the main properties of the  model discussed so far. 
The Hamiltonian,  Eq.(\ref{m8}),  describes the dynamics of 
 a massless Dirac  fermion in the  region of Rindler spacetime bounded
by the hyperbola $\rho = \ell_1$.  The reflection of the wave function at this boundary
is characterized by a parameter  $\vartheta$, which is real for a self-adjoint  Hamiltonian.
At the positions $\ell_{n >1}$
the wave function  is discontinuous due to the presence of delta function potentials
characterized by the reflections amplitudes $\varrho_n$,  that  provide
the matching conditions of the wave function at those sites.  
An  eigenfunction $\chi$,  with eigenvalue $E$,   has a simple expression, Eq.(\ref{m17}), 
in terms of  the amplitudes $A_{n, \pm}$, that are related by the transfer matrix  $T_n$ (\ref{m21}). 
The norm of $\chi$ is given by the sum of the squared length  of the vectors ${\bf A}_n$, weighted
with a factor that depends on the positions $\ell_n$, Eq.(\ref{m22}).  We  introduce 
an scale factor $\varepsilon$  in the parameters $\varrho_n$, that allows us  to study  the
limit $\varepsilon \rightarrow 0$, where the mirrors become semi transparent. In this way we found 
an  ansatz  for the parameters $\ell_n$ and $\varrho_n$ that heuristically led to an individual
spectral realization of the {\em zeros} by fine tuning the parameter $\vartheta$. 

\subsubsection{Normalizable eigenstates} 

 Under  the choice  $\ell_n = n^{ 1/2}$, Eq.(\ref{m22}) becomes 
\barray 
 || \chi ||^2  = 
\frac{1}{2} \sum_{n=1}^\infty \log \left( 1 + \frac{1}{n} \right)   \, \langle {\bf A}_n | {\bf A}_n \rangle \, . 
\label{r1}  
\earray 
This series can be replaced by 
%
\barray 
 || \chi ||_c^2  \equiv 
 \sum_{n=1}^\infty  \frac{1}{n}    \, \langle {\bf A}_n | {\bf A}_n \rangle \, , 
\label{r2}  
\earray 
which is convergent if and only if (\ref{r1}) is convergent.   The vectors  ${\bf A}_n$ 
are obtained by acting on   ${\bf A}_1(\vartheta)$  with the transfer matrices $T_n$ (see Eq.(\ref{m23})).
These matrices have unit determinant and can be written as the exponential of traceless hermitean 
matrices, that is, 
\beq
T_n  = e^{ \tau_n}, \qquad 
\tau_n = \left( \begin{array}{cc}
0 & r_n \ell_n^{ - 2 i E} \\
r_n^*  \ell_n^{ 2 i E} & 0 \\
\end{array}
\right) , \qquad \forall  E \in \Rmath \, , 
\label{r3}
\eeq
where taking  $|\varrho_n | < 1$, 
\beq
r_n = \frac{ \varrho_n}{ |\varrho_n|} \log \frac{ 1 +  |\varrho_n|}{ 1-  |\varrho_n|} ,  \qquad \varrho_n = \frac{ r_n}{|r_n|} \tanh \frac{ |r_n|}{2} \, . 
\label{r4}
\eeq
To derive Eq.(\ref{r3})   we used the relation
\beq
{\rm exp} 
 \left( \begin{array}{cc}
0 & a  \\
b   & 0 \\
\end{array}
\right) = 
 \left( \begin{array}{cc}
\cosh( \sqrt{ a b} )  & \frac{a}{ \sqrt{ a b  }} \sinh ( \sqrt{ a b} )   \\
\frac{b}{ \sqrt{ a b  }} \sinh  ( \sqrt{ a b} ) & \cosh( \sqrt{ a b} ) \\
\end{array}
\right) , \quad \forall a, b \in \Cmath - \{   0 \} \, .  
\label{r5}
\eeq
If $|\varrho_n| << 1$ one gets   $r_n \simeq  2 \varrho_n$, hence in that limit both parameters give  the same result.
Using Eq.(\ref{r3}), the recursion relation (\ref{m23}) reads 
\beq
| {\bf A}_{k}  \rangle  =   e^{ - \tau_k}    e^{ - \tau_{k-1}}    \dots  e^{ - \tau_2}   | {\bf A}_{1}  \rangle, \quad k \geq 2. 
\label{r6}
\eeq

\subsubsection{The Magnus expansion} 

It is rather difficult to find an analytic expression of the product of matrices of Eq.(\ref{r6}).  
However, we  can estimate it  replacing $r_n$ by $\varepsilon r_n$, and 
taking   the limit  $\varepsilon \rightarrow 0$. Under this replacement Eq.(\ref{r6}) becomes 
\beq
| {\bf A}_{k}  \rangle    = e^{ - \varepsilon  \tau_k}    e^{ -  \varepsilon  \tau_{k-1}}    \dots  e^{ -  \varepsilon \tau_2}    | {\bf A}_{1}  \rangle \quad  (k  \geq 2).  
\label{r7}
\eeq
The product of exponentials of matrices  can be expressed as  the   exponential of a  matrix given by the Magnus expansion  \cite{Magnus} 
\beq
 e^{ - \varepsilon  \tau_k}    e^{ -  \varepsilon  \tau_{k-1}}    \dots  e^{ -  \varepsilon \tau_2}  = {\rm exp} \;
 \left( - \varepsilon \sum_{n=2}^n \tau_n - \frac{\varepsilon^2}{2}  \sum_{n_1 > n_2 =2}^k [ \tau_{n_1}, \tau_{n_2} ]  +  
O(\varepsilon^3)  \right) 
 \quad  (k  \geq 2).  
\label{r8}
\eeq
In the limit $\varepsilon \rightarrow 0$ we  truncate this expression to the  term of order $\varepsilon$, 
\beq
 e^{ - \varepsilon  \tau_k}    e^{ -  \varepsilon  \tau_{k-1}}    \dots  e^{ -  \varepsilon \tau_2}  \simeq {\rm exp} \;
 \left( 
 \begin{array}{cc}
 0 & - \varepsilon \sum_{n=2}^k r_n \, \ell_n^{ - 2 i E} \\
- \varepsilon \sum_{n=2}^k r_n^*  \, \ell_n^{  2 i E}  & 0 \\
\end{array} 
\right) \simeq {\rm exp} \
  \left( 
 \begin{array}{cc}
 0 & - \varepsilon M_z(k)  \\
- \varepsilon M^*_z(k)  & 0 \\
\end{array} 
\right) \, , 
\label{r9}
\eeq
which  using  (\ref{m27})) 
\beq
r_n = \frac{ \mu(n)}{n^{1/2}}
\label{r9b}
\eeq
gives 
\beq
M_z(k) = 1+ \sum_{n=2}^k r_n \, \ell_n^{ - 2 i E} = \sum_{n=1}^k \frac{\mu(n)}{n^z} , \qquad z = \frac{1}{2} + i E   \, . 
\label{r10}
\eeq
We have added the constant  1 to $M_z(k)$,  that does not affect the results in the limit $\varepsilon \rightarrow 0$  . 
Using Eqs.(\ref{r5}), (\ref{r7}) and  (\ref{r9}) we obtain 
\barray 
| {\bf A}_{n}  \rangle  & \simeq & 
{\rm exp} \
  \left( 
 \begin{array}{cc}
 0 & - \varepsilon M_z(n)  \\
- \varepsilon M^*_z(n)  & 0 \\
\end{array} 
\right)  \left( \begin{array}{c}
1 \\
e^{i \vartheta} \\
\end{array} 
\right) \label{r11} \\  
& = & 
  \left( 
 \begin{array}{cc}
 \cosh( |  \varepsilon  M_z(n) | )   & -  \frac{  \varepsilon M_z(n) }{  |  \varepsilon M_z(n) | }    \sinh ( |  \varepsilon  M_z(n)|)   \\
-  \frac{  \varepsilon M_z^*(n) }{  |  \varepsilon M_z(n) | }    \sinh ( |  \varepsilon  M_z(n)|  &   \cosh( |  \varepsilon  M_z(n) | )  \\
\end{array} 
\right)  \left( \begin{array}{c}
1 \\
e^{i \vartheta} \\
\end{array} 
\right)  
\nonumber \\
& = & 
  \left( 
 \begin{array}{cc}
 \cosh( |  \varepsilon  M_z(n) | )   & -  e^{ - i \Phi_z(n) }   \sinh ( |  \varepsilon  M_z(n)|)   \\
- e^{  i \Phi_z(n) }   \sinh ( |  \varepsilon  M_z(n)|  &   \cosh( |  \varepsilon  M_z(n) | )  \\
\end{array} 
\right)  \left( \begin{array}{c}
1 \\
e^{i \vartheta} \\
\end{array} 
\right)  \nonumber   \\ 
& \simeq & 
 \left( \begin{array}{c}
 e^{  \frac{i}{2}  (\vartheta - \Phi_z(n) )}  \left[ 
e^{ - | \varepsilon  M_z(n) |}   \cos( \frac{1}{2}  (\vartheta - \Phi_z(n) )  -  i  e^{ | \varepsilon  M_z(n) | }  \sin( \frac{1}{2}  (\vartheta - \Phi_z(n) ) \right]    
  \\  \\
   e^{ \frac{i}{2}  (\vartheta + \Phi_z(n) )} \left[ 
e^{ - | \varepsilon  M_z(n) |}   \cos( \frac{1}{2}  (\vartheta - \Phi_z(n) )  +  i  e^{  | \varepsilon  M_z(n) | }  \sin( \frac{1}{2}  (\vartheta - \Phi_z(n) ) 
\right]    \\
\end{array}
\right) 
\quad  (n  \geq 2) \, , 
\nonumber 
\earray 
where  $\Phi_z(n)$ is the  phase 
\beq
e^{ - i \Phi_z(n) } = \frac{ M_z(n)}{ |M_z(n) |} \, . 
\label{r13}
\eeq
From (\ref{r11}) follows  an estimate of the norm  (\ref{r2}) 
\barray 
 || \chi ||_c^2  \simeq  {\cal N}_z(\varepsilon) \equiv 
  \sum_{n=1}^\infty  \frac{1}{n}   \left[ 
   e^{ -  2 | \varepsilon  M_z(n) |}  ( 1+   \cos   (\vartheta - \Phi_z(n) )   
   +  e^{ 2  | \varepsilon  M_z(n) | }  ( 1 - \cos  (\vartheta - \Phi_z(n) )   \right] \, ,
\label{r12}  
\earray
whose  convergence depends on  the asymptotic behaviour of $M_z(n)$ and $\Phi_z(n)$. 
${\cal N}_z(\varepsilon)$ has the lower bound 
\beq
{\cal N}_z(\varepsilon)  \geq    \sum_{n=1}^\infty  \frac{2}{n}   e^{ -  2 | \varepsilon  M_z(n) |}  \, , 
\label{r13b}
\eeq 
that follows from the inequality 
\beq
a ( 1 + b) + \frac{1}{a} ( 1 - b) \geq 2 a, \qquad  a \in(0,1], \quad b \in [-1, 1] \, . 
\label{r14}
\eeq
If $|M_z(n)|$ is a bounded the norm is infinite, 
\beq
{\rm if} \; |M_z(n) |  < C, \; \; \forall n \Longrightarrow
{\cal N}_z(\varepsilon)  \geq    \sum_{n=1}^\infty  \frac{2}{n}   e^{ -  2 | \varepsilon |  C }   = \infty  \, . 
\label{r15}
\eeq 
This case corresponds in general to eigenstates belonging to the continuum.  
Eigenstates with finite norm require  $|M_z(n)|$ to be unbounded.   Notice that 
${\cal N}_z(\varepsilon)$
is the sum of two series with non negative  terms. The convergence of the first summand  in (\ref{r12})
is guaranteed if 
\beq
 \sum_{n=1}^\infty  \frac{1}{n}   e^{ -  2 | \varepsilon M_z(n)|}   <  \infty  \, , 
\label{r16}
\eeq
which occurs if $|M_z(n)|$ diverges sufficiently fast with $n$. 
The convergence of the second summand  in (\ref{r12}) requires 
$\Phi_z(n)$ to have  a limit when $n \rightarrow \infty$, and to choose the parameter $\vartheta$
such that 
\beq
\lim_{n \rightarrow \infty} \Phi_z(n)  = \vartheta \, .
\label{r18}
\eeq 
Moreover $1 - \cos  (\vartheta - \Phi_z(n))$ must approach 0 sufficiently fast in order 
to compensate the factor $\frac{1}{n} e^{ 2 \varepsilon |M_z(n)|}$.  
We now pass to analyze the  latter  conditions in detail.

\subsubsection{Perron formula}

%
%
%

\begin{figure}[h!]
\begin{center}
\includegraphics[height= 5.0 cm]{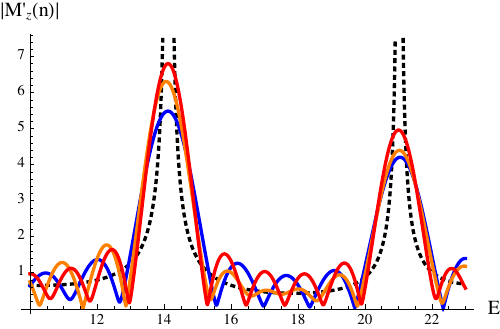}
\end{center}
\caption{Plot of $|M'_z(n)|$ defined in  Eq.(\ref{r48}), 
 for $E \in (10,23)$  and  
$n=50, 100, 150$  (blue, orange, red curves) and $1/|\zeta(1/2 + i E)|$ (black dotted line). 
Observe the increase with $n$  at  $E=14.13 \dots$ and $E=21.02 \dots$ which are the first two {\em zeros} of $\zeta$. 
} 
\label{Gn}
\end{figure}

\begin{figure}[t]
\begin{center}
\includegraphics[height= 4.0 cm]{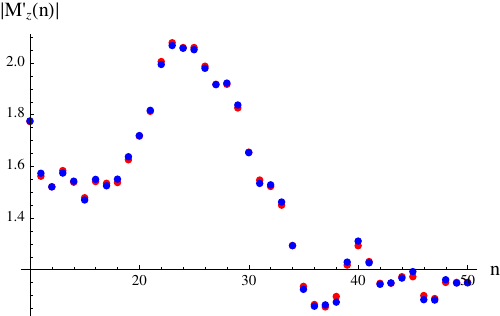}
\hspace{0.2cm}
\includegraphics[height= 4.0 cm]{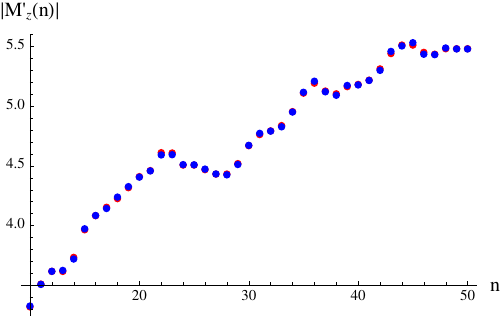}
\end{center}
\caption{Plot of $|M'_z(n)|$ for $n=10, \dots, 50$ and 
 $E=20$  (left)   and $E= 14.13..$ (right). In red the values obtained doing  the sum in Eq.(\ref{r48}).
 In blue the sum of Eq.(\ref{r541}) for $E=20$ and Eq.(\ref{r542}) for $E= 14.13..$, including
 the first 100 Riemann zeros, and 20 trivial zeros. Observe the accuracy of the approximation. 
 The slow increase in the latter
 plot is due to the factor $\log n$  in Eq.(\ref{r542}). 
} 
\label{Gn20}
\end{figure}

Let us  define the  function 
\beq
M'_z(x) \equiv 
{\sum_{1 \leq n \leq x}}'  \;  \frac{  \mu(n)}{n^z}  \, , \qquad z = \frac{1}{2} + i E, \quad E \in \Rmath \, , 
\label{r48}
\eeq
where  ${\sum'_{1 \leq n \leq x}} $ means that the last term  in the sum is  multiplied by 1/2 
when $x$ is an integer. 
Fig. \ref{Gn} shows $M'_z(n)$ as a function of $E$ for  several values of $n$.
Observe that  $|M'_z(n)|$ increases with $n$  when $E$ is a {\em zero}. 
We shall derive  below this behavior. 
To  compute  $M'_z(x)$    we  use  Perron's  formula  \cite{Apostol}
\beq
M'_z(x) = \lim_{T \rightarrow \infty} \int_{c- i T}^{c + i T} \frac{d s }{ 2 \pi i}  \frac{1}{\zeta(s+z)}  \frac{x^s}{s}, 
\qquad c >  \frac{1}{2} \,  , 
\label{r48d}
\eeq
where we have  used   that  ${\rm Re} \, z = 1/2$. 
The integral (\ref{r48d}) can be done by residue calculus \cite{S14}
\beq
 \lim_{T \rightarrow \infty} \int_{c- i T}^{c + i T} \frac{d s }{ 2 \pi i} F(s) 
 = \sum_{ {\rm Re} \, s_j < c } {\rm Res}_{s_j}   \,   F(s), \quad F(s) =   \frac{1}{\zeta(s+z)}  \frac{x^s}{s}, 
\label{r49}
\eeq
where the sum runs  over the  poles $s_j$  of $F(s)$ located to the left of the line of integration  ${\rm Re} \; s =c$, that is 
${\rm Re} \, s_j < c$.  
The poles of $F(s)$ come  from  the zeros of $s \zeta(s+z)$.  The pole at $s=0$ can be simple, or multiple,  depending
on the values of $\zeta(z)$ and its derivatives. The remaining poles of $F(s)$ 
come from the zeros of  $\zeta(s+z)$, say  $s_j +z = \rho_j$, and they  lie  to  the left   of the integration line, 
because the  trivial and non trivial {\em zeros} of $\zeta$, satisfy ${\rm Re}  \;  \rho_j < 1$, that is 
\beq
{\rm Re} \; s_j = {\rm Re} ( \rho_j - z) = {\rm Re}  \;  \rho_j - \frac{1}{2} < \frac{1}{2}   < c  \, . 
\label{r50}
\eeq
To compute the  residues of Eq.(\ref{r49}) 
we consider the cases: $s=0$,  $s_j + z$ a  trivial zero   of $\zeta$  and  $s_j + z$ a non trivial zero of $\zeta$:  

\begin{itemize}

\item $s=0$. Let $m \geq 0$ be the lowest integer such that $\zeta^{(m)}(z) = d^m \zeta(z)/dz^m  \neq 0$. Then $F(s)$ 
has a pole of order $m+1$ at $s=0$ with residue \cite{core}  
\beq
{\rm Res}_{s=0}   \,   F(s) = \left\{
\begin{array}{ll}
1/\zeta(z)   & {\rm if} \, \zeta(z) \neq 0,  \\ 
\log x/\zeta'(z) - \frac{1}{2}  \zeta''(z)/(\zeta'(z))^2  & {\rm if} \, \zeta(z) = 0, \zeta'(z) \neq   0. \\ 
\vdots & \vdots \\ 
(\log x)^m/\zeta^{(m)}(z) + O((\log x)^{m-1} )  & {\rm if} \, \zeta(z) =  \dots =   \zeta^{(m-1)}(z) =  0 ,\zeta^{(m)}(z)   \neq    0. \\ 
\end{array}
\right. 
\label{r51}
\eeq

\item  $s_n = - 2 n - z \; (n=1, 2, \dots)$, where $F(s)$ has a simple pole due to the trivial {\em zeros} $- 2n$ of $\zeta$. 
\beq
{\rm Res}_{s=- 2 n - z}   \,   F(s) =  \frac{ x^{ - 2 n - z} } { - ( 2 n + z) \zeta'( - 2 n )}, \qquad n =1, 2, \dots, \infty .  
\label{r52}
\eeq

\item $s_j = \rho_j - z \neq 0$, then  $F(s)$ has a  pole due to the non trivial zero $\rho_j$ of $\zeta$
%
 \beq
{\rm Res}_{s= s_j }   \,   F(s) = 
\left\{  \begin{array}{ll} 
 \frac{ x^{ \rho_j  - z} } { ( \rho_j - z) \zeta'( \rho_j )}, & {\rm if} \; \zeta(\rho_j) =0, \zeta'(\rho_j) \neq 0 \\ \\
  \frac{ m (\ln x)^{m-1}x^{ \rho_j  - z} } { ( \rho_j - z) \zeta^{(m)}( \rho_j )} + {O} ( (\ln x)^{m-2}) , & {\rm if} \; \zeta(\rho_j) =
   \dots=\zeta^{(m-1)} (\rho_j) =0, \zeta^{(m)}(\rho_j)   \neq 0, \; m \geq 2 \\
\end{array}
\right. 
\label{r53}
\eeq
%
%
\end{itemize}
To make further progress we shall make the assumption that all the Riemann zeros are simple,
a statement which is not known to hold. The eventual case where there is a {\em zero} with double multiplicity
will be considered elsewhere.  In the former  situation we are led to consider  only two cases
depending on whether $z$ is, or is not,  a simple {\em zero}  of $\zeta$. Collecting terms we get 
\barray 
M_z(x) 
&  = &  \frac{1}{\zeta(z)} + 
\sum_{  \rho_j } 
 \frac{ x^{ \rho_j  - z} } { ( \rho_j - z) \zeta'( \rho_j )}  +  \sum_{n=1}^{\infty}  \frac{ x^{ - 2 n - z} } {  - ( 2 n + z) \zeta'( - 2 n )},  \quad  {\rm if} \;   \zeta(z) \neq 0 \, ,  
\label{r541}  \\ 
M_z(x)  
&  = &  \frac{\log x }{\zeta'(z)} - \frac{ \zeta''(z)}{ 2 (\zeta'(z))^2} + 
\sum_{ \rho_j  \neq z}  \frac{ x^{ \rho_j  - z} } { ( \rho_j - z) \zeta'( \rho_j )} +   \sum_{n=1}^{\infty}  \frac{ x^{ - 2 n - z} } { - ( 2 n + z) \zeta'( - 2 n )},   \; \; 
{\rm if} \; \zeta(z) =0, \zeta'(z) \neq 0 \, . 
\label{r542} 
\earray 
where the sum  $\sum_{\rho_j}$ runs  over the non trivial zeros of $\zeta$. 
These equations are verified numerically in Fig.\ref{Gn20}. 
%
%
 The last term in  these equations, that comes from the trivial
{\em zeros},  converges quickly and is finite for all $x$ due to the exponential increase of $\zeta'(-2 n)$  \cite{CMS} 
\beq
\zeta'(-2n) = \frac{ (-1)^n \zeta(2 n + 1) (2 n)! }{ 2^{ 2 n +1} \pi^{2 n}} \; \;  \stackrel{n \rightarrow \infty}{\longrightarrow} \; 
 (-1)^n \sqrt{ \pi n} \left( \frac{ n}{ e \pi} \right)^{ 2 n} \, . 
\label{r55b}
\eeq


%

%
We do not know an estimation of the term depending on the sum over the non trivial zeros. 
If the Riemann hypothesis is true the term $x^{\rho_j - z}$ will oscillate as a function of $x$.  We expect that
for $\zeta(z) \neq 0$, $|M_z(x)|$  will not yield a finite norm such that 
the corresponding eigenstate  will not belong to the  discrete spectrum. When $\zeta(z))=0, \zeta'(z)\neq 0$, 
we shall make the approximation 

\barray 
M_z(x)  
&  \rightarrow   &  \frac{\log x }{\zeta'(z)} 
 \qquad x \rightarrow \infty  
\label{r542} 
\earray 
There could exists a  finite part in this expression, in particular the term 
$\frac{ \zeta''(z)}{ 2 (\zeta'(z))^2}$, however the term involving the sum over the Riemann zeros
may give additional contributions. Using that $\zeta( 1/2 + i E) = e^{- i \theta(E)} Z(E)$ we find

\begin{figure}[h!]
\begin{center}
{\includegraphics[height= 2.5 cm]{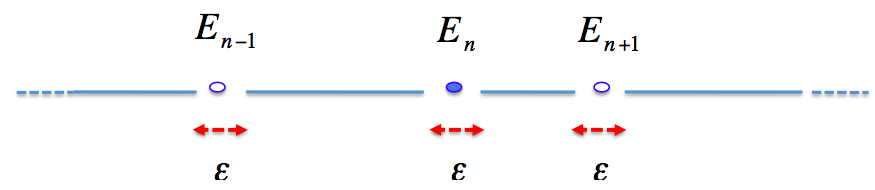}
}
\end{center}
\caption{Graphical  representation of the spectrum of the model. It is expected to consists 
of an infinite number of bands separated by forbidden regions of width proportional to   $\varepsilon$. 
The latter regions may contain a {\em zero} $E_n$ if the phase $\vartheta$ is chosen according to Eq.(\ref{r64}). 
Otherwise the {\em zeros} will be missing in the spectrum, that are represented  by the  points
$E_{n-1}$ and $E_{n+1}$.   
} 
\label{Spectrum}
\end{figure}

%
%
%
\barray 
M_x(z)&  \rightarrow   & 
 \frac{ i  \,  e^{ i \theta(E)} \log x  }{ Z'(E)}  \, \quad  {\rm as} \; \;  x \rightarrow \infty, \neq 0, \, 
 \label{r62} 
\earray 
hence $\Phi_z(n)$, given in Eq.(\ref{r13}),  behaves as
\beq
e^{- i \Phi_z(n)} \rightarrow  i  \,  e^{ i \theta(E)}  {\rm sign} \,  Z'(E)   \quad  {\rm as} \; \;  n \rightarrow \infty \, , 
\label{r63}
\eeq
which has a well defined asymptotic limit. We shall then  choose $\vartheta$ according to eq(\ref{r18}) 
namely
\beq
\vartheta = \lim_{n \rightarrow \infty} \Phi_z(n) = - \left( \theta(E)  + \frac{\pi}{2}  {\rm sign } \,  Z'(E) \right) \, ,
\label{r64}
\eeq
that provides a necessary condition for the convergence of the norm. It remains to show
that eq.(\ref{r64}) is also sufficient  but this requires the knowledge of  the next to leading correction to (\ref{r62}). 
Notice that  $\vartheta$ depends on $\theta(E)$ and  the sign of $Z'(E)$, a feature
that is not left fixed  in eq.(\ref{m29b}).  The norm (\ref{r12}) then becomes 
\barray 
 || \chi ||_c^2  \simeq 
 \sum_{n=1}^\infty  \frac{2}{n}  e^{ -  2 \varepsilon \log n /|Z'(E)|} = 2 \zeta \left( 1 + \frac{ 2 \varepsilon}{ |Z'(E)|} \right)   < \infty \, ,
\label{r65}  
\earray
that is finite for all $\varepsilon >0$. This result  indicates   that a {\em zero} of the zeta function 
gives a normalizable state, in agreement with  heuristic derivation proposed 
in  the previous section, but there are some differences. First of all, the eigenvalue $E$ does not need
to be expanded in series of $\varepsilon$. It is taken to be 
a {\em zero} of $\zeta$ from the beginning. This choice  generates  the $\log x$ term in Eq.(\ref{r542})
and is responsible for  the finiteness of the norm after the appropriate choice of the phase (\ref{r64}), that also
differs from the heuristic value (\ref{m29b}).  On the other hand, if $\vartheta$ does not satisfy Eq.(\ref{r64}),
then the  norm of the state will diverge badly  and so the {\em zero}  $E$  will be missing in the spectrum. 
Finally, if $E$ is not a {\em zero}, we expect that the state will belong generically to the continuum. 
Fig. \ref{Spectrum}  shows the expected spectrum of the model, which recalls Connes's  scenario
of missing spectral lines,  except that in our case, one can pick up a {\rm zero} at a time by tuning $\vartheta$.

 If the RH is false there will be at least  four   {\em zeros} outside  the critical line, say 
 $\rho_c = \sigma_c + i E_c, \bar{\rho}_c = \sigma_c - i E_c,  1 - \rho_c$ and $1 - \bar{\rho}_c$,  
 with $\sigma_c > \frac{1}{2}, E_c \in  \Rmath_+$. 
 We shall  choose the highest value of $\sigma_c$. 
The asymptotic behavior of $M_z(x)$  will be given by the  {\em zeros} located to the right of the critical line, 
\barray 
M_z(x) 
&  \rightarrow  &  
 \frac{ x^{ \rho_c  - z} } { ( \rho_c - z) \zeta'( \rho_c )}   +    \frac{ x^{ \bar{\rho}_c  - z} } { ( \bar{\rho}_c - z) \zeta'(  \bar{\rho}_c )}
\, \quad  {\rm as} \; \;  x \rightarrow \infty \, . 
\label{r66}
\earray 
To simplify the discussion let us choose $E \gg E_c$, which yields  the approximation
\barray 
M_z(x) 
&  \rightarrow  &  \frac{2  i  \, x^{ \sigma_c - 1/2 - i E}  }{ E |\zeta'(\rho_c)|} \cos( E_c \log x - \phi_c) 
\, \quad  {\rm as} \; \;  x \rightarrow \infty \, , 
\label{r67}
\earray 
where $e^{i \phi_c} = \zeta'(\rho_c)/ |  \zeta'(\rho_c)|$. The phase $\Phi_z(n)$ is given by  Eq.(\ref{r13}) 

\beq
\Phi_z(n)  \rightarrow   E \log n  -  \frac{ \pi}{2}  {\rm sign} \;  ( \cos( E_c \log n  - \phi_c)  ) 
\, \quad  {\rm as} \; \;  n \rightarrow \infty \, . 
\label{68}
\eeq
Correspondingly,  the norm (\ref{r12})   diverges so badly,   $ \propto \sum_n \frac{1}{n}  {\rm exp} (C  n^{\sigma_c - 1/2}) \dots$,
for any value of $\vartheta$,  that the state will not be normalizable even using  Dirac delta functions. 
This result occurs  for all  eigenenergies $E$.  Therefore  the Hamiltonian will not admit a spectral decomposition, 
but this is impossible because it is a well defined self-adjoint operator. We  conclude that a  {\em zero} 
outside the critical line does not exist which provides an 
argument likely to be persuasive to physicists for the truth of the Riemann
hypothesis. 


\section{The Riemann  interferometer}

\begin{figure}[h!]
\begin{center}
{\includegraphics[height= 4.5 cm]{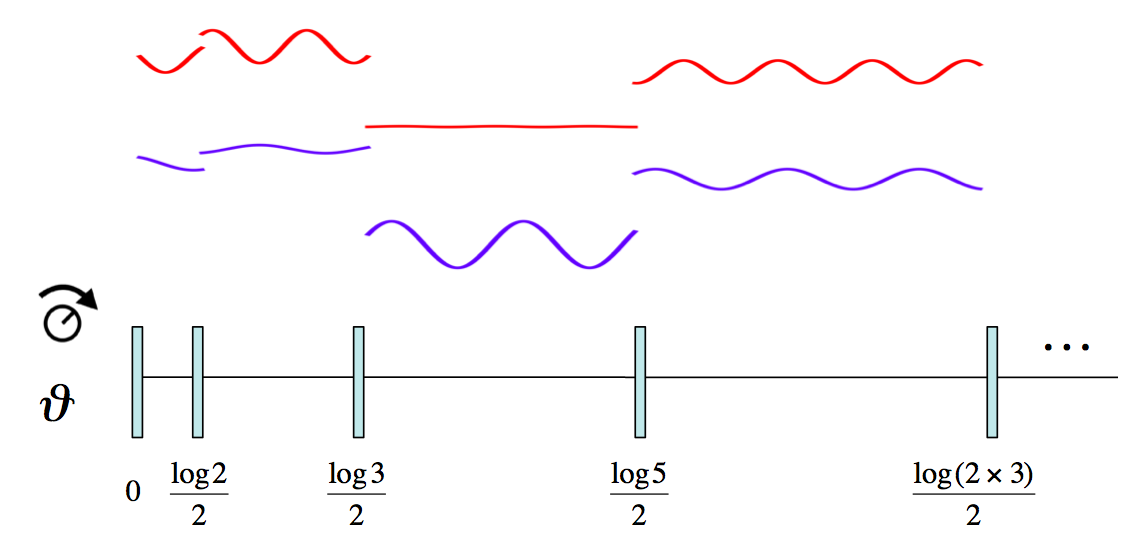}
}
\end{center}
\caption{Graphical representation of the array of mirrors in Minkowski  space that  reproduce the Riemann zeros. 
The phase at the boundary $\vartheta$ has to be chosen according to Eq.(\ref{r64}) in order that $E$ is an eigenvalue
of the Hamiltonian. Recall Fig. \ref{tune-zero}. Between the mirrors the wave functions are plane waves. 
} 
\label{ARRAY2}
\end{figure}

The model considered  in the previous  sections looks at first glance  quite  difficult to simulate. 
We  shall next show that  this model  is equivalent to another one that can be implemented  in the Lab. 
We shall call this system the Riemann  interferometer. 
 The basic idea can be illustrated with the   mapping between the quantum $xp$  hamiltonian 
and the momentum operator $\hat{p}$. Let us make the change of coordinates  $x = \log \rho$
and relate the wave functions in both coordinates,  $\phi(x)$ and   $\psi(\rho)$,  as follows
\beq
\phi(x) = \left( \frac{d\rho}{dx} \right)^{ 1/2} \psi(\rho)  = e^{ x/2} \psi(e^x)  \, . 
\label{e1}
\eeq
An eigenstate of the  Hamiltonian $(\rho  \,  \hat{p}_\rho + \hat{p}_\rho \,  \rho )/2$,  with eigenvalue $E$, is mapped by Eq.(\ref{e1}) 
into  an eigenstate of the momentum 
operator $\hat{p}_x = - i \partial_x$ with the same eigenvalue, 
\beq
\psi(\rho) = \frac{1}{ \rho^{ 1/2 - i E}} \Longrightarrow \phi(x) = e^{ i E x}  \, . 
\label{e2}
\eeq
This shows that the energy $E$ can  be seen as momentum.  For a relativistic massless fermion,
this is always the case. 
The measure that defines the scalar product of the corresponding Hilbert spaces are  one-to-one related
\beq
\int_\ell^\infty d\rho \, \psi^*_1(\rho) \psi_2(\rho) = \int_{\log \ell}^\infty dx \, \phi^*_1(x) \phi_2(x) \, .
\label{e3}
\eeq
The operator  $(\rho  \,  \hat{p}_\rho + \hat{p}_\rho \,  \rho )/2$ is self-adjoint in the interval $(0, \infty)$ 
but  not in the interval $(1, \infty)$, just like $\hat{p}_x$ is self-adjoint in the real line $(- \infty, \infty)$ 
but not in the  half line $(0, \infty)$ \cite{GP90,S07a}. The former case corresponds to the value $\ell=0$
and the latter  one to  $\ell=1$ in Eq.(\ref{e3}).   Let us now consider the Dirac Hamiltonian in the Rindler variable 
 $\rho$,  given in Eq.(\ref{m8}). It becomes in the $x$ variable 
\beq
H = \left(
\begin{array}{cc} 
 - i \partial_x  & 0 \\
 0 & i \partial_x \\
 \end{array}
 \right) \, . 
 \label{e4}
 \eeq
 Unlike $\hat{p}_x$, this Hamiltonian is  self-adjoint in the interval $x \in (\log \ell_1, \infty)$.
We choose for convenience $\ell_1 =1$. The moving mirrors located at  $\rho = \ell_n$ are now placed
at the positions $x = d_n$, with $d_n = \log \ell_n$, so for $\ell_n = \sqrt{n}$, we get 
\beq
d_n = \frac{1}{2} \log n \, , 
\label{e6}
\eeq
where $n$ are square free integers  and the reflection coefficients are given by $r_n= \mu(n)/\sqrt{n}$.
Fig. \ref{ARRAY2} shows the array of mirrors satisfying Eq.(\ref{e6}). One can easily  generalize this interferometer 
to provide a spectral realization of the {\em zeros} of Dirichlet $L$-functions, by  changing the reflection
coefficients $r_n$, 
\beq
L_\chi(s) = \sum_{n=1}^\infty \frac{ \chi(n)}{ n^s} \longrightarrow r_n = \frac{ \mu(n) \, \chi(n)}{n^{1/2}} \, , 
\label{e7}
\eeq
where $\chi(n)$ is the Dirichlet character associated to the $L$-function. It would be interesting to replace
the massless fermions by massless bosons, say photons and study what kind of Riemann interferometer 
arise.

\section{Dirac  models for a class of  modified $\zeta$  and  $L$ functions} 

Grosswald and Schnitzer proved in 1978  two  very surprising theorems that we shall use below
to generalize the construction done in previous sections.  Let us first consider a set on integers $q_n$
satisfying the conditions 
\beq
p_n \leq q_n \leq p_{n+1}, \qquad n=1, \dots,  \infty  \, ,
\label{gs1}
\eeq
where $p_n$ is the $n^{\rm th}$ prime number. With these numbers define the infinite product
\beq
\zeta^*(s) = \prod_{n=1}^\infty ( 1- q_n^{-s})^{-1} \, .
\label{gs2}
\eeq 
One then has \cite{GS78}: 
\vspace{0.1cm} 

{\bf Theorem 1} : This function is holomorphic for $\sigma = {\rm Re} \,  s > 1$ and has the following properties: 
\barray
i) &  &  \zeta^*(s)  \neq 0, \; \;  {\rm for} \; \; \sigma > 1,  \nonumber   \\
ii) & & \zeta^*(s)  \; {\rm has} \; {\rm a } \; \; {\rm meromorphic} \; \; {\rm extension}\; \; {\rm to} \; \; \sigma > 0,  \nonumber  \\
iii) & & {\rm in} \; \; \sigma > 0,  \; \; 
\zeta^*(s)  \; {\rm has} \; {\rm a } \; \; {\rm simple} \; \; {\rm pole}\; \; {\rm at}  \; \; s=1 \; \; {\rm with } \; \; {\rm residue} \; 
\; r ,  \;  1/2  \leq r \leq 1,  \nonumber  \\
iii) & & {\rm in} \; \; \sigma > 0,  \; \; 
 \zeta^*(s)  \; {\rm has} \; {\rm the  } \; \; {\rm same} \; \; {\rm zeros}\; \; {\rm as} \; \;  \zeta(s) \; \; {\rm with } \; {\rm the} 
\; \; {\rm same}  \; \; {\rm multiplicity}.  \nonumber 
\earray 
This theorem means that the relation between prime numbers and Riemann zeros via the zeta function 
is less rigid that one may have though. We shall use this freedom to associate 
a Hamiltonian to every series satisfying (\ref{gs1}).  Let us first write  the inverse of (\ref{gs2}) as 
\beq
\frac{1}{\zeta^*(s)}  = \sum_{n=1}^\infty \frac{ \mu^*(n)}{ n^s} \, , \qquad \mu^*(n) = n_{\rm even} - n_{\rm odd}, 
\label{gs3}
\eeq
where $n_{\rm even} (n_{\rm odd})$ is  the number of times $n$ can be written as the  product
of an even (odd)  number of $q_i$ numbers in the series (\ref{gs1}). 
An example of a series satisfying  (\ref{gs1}) is  
%
%
\beq
2, 4, 6, 8, 12, \dots q_n = p_n +1, \dots 
\label{gs4}
\eeq 
for which we have 
\beq
\frac{1}{\zeta^*(s)} = 1 - \frac{1}{2^s}  -  \frac{2}{ (2^6  \cdot 3)^s}  +   \frac{2}{(2^3  \cdot 3)^s}  +  
\frac{1}{ (2^8 \cdot  3)^s} - \frac{1}{2^{2s}} - \frac{1}{(2 \cdot 3)^{s}} + \dots  \, .
\label{gs5}
\eeq
Notice that $\mu^*(2^6 \cdot 3) = -2$ because $2^6 \cdot 3 = 4 \cdot 6 \cdot 8 = 2 \cdot 8 \cdot 12$. 
Obviously $\mu^*(n) = \mu(n)$ if $q_n= p_n, \; \; \forall n$. Using eq.(\ref{gs3}) 
we define a massless Dirac  model with reflection coefficients  (recall eq.(\ref{r9b}) ) 
\beq
r_n  = \frac{ \mu^*(n)}{n^{1/2}}, \quad n > 1  \, . 
\label{gs6}
\eeq 
Hence, by the arguments given in section XII and theorem 1, 
we shall find the Riemann zeros in the spectrum of the Hamiltonian $H_\vartheta$ 
 by tuning the parameter $\vartheta$ in the limit $\varepsilon \rightarrow 0$. 
 
 The second theorem in reference \cite{GS78} is an extension  of theorem 1
 to  Dirichlet $L$-functions $L(s) = \prod_n (1 - \chi(n) n^{-s})^{-1}$, where 
 $\chi$ is a character modulo $k$. The series (\ref{gs1}) is replaced by 
\beq
p_n \leq q_n \leq p_{n} + K,  \qquad p_n = q_n \; {\rm mod} \; k
\label{gs7}
\eeq
where  $K \geq k$.  The modified character is defined as 
\beq
L^*(s) = \prod_{n=1}^\infty ( 1- \chi(q_n) q_n^{-s})^{-1} \, ,
\label{gs8}
\eeq 
 that can be extended to the region $\sigma >0$, with the same zeros (and multiplicities) as $L(s)$. 
 In this case too, we can construct a Dirac model with reflection coefficients (recall eq.(\ref{e7}))
\beq
r_n  = \frac{ \chi(n) \mu^*(n)}{n^{1/2}}, \quad n > 1  \, . 
\label{gs9}
\eeq 
whose associated Hamiltonian $H_\vartheta$ contains the zeros of $L(s)$ by varying $\vartheta$.  
Theorem 2 of \cite{GS78} was mentioned by LeClair and Mussardo in \cite{ML18b}  as a support to  their   approach
to the Generalized Riemann hypothesis  based on random walks and 
the  Lemke Oliver-Soundararajan conjecture on the distribution of pairs of
residues on consecutive primes \cite{LOS16} (for other statistical properties of the prime numbers see \cite{SK17}). 
 It will be worth to investigate
if there is a relation between our  approach and the one proposed  in \cite{ML18a,ML18b}.


\section{Conclusions}

In this paper we have reviewed the spectral approach to the RH that started with the Berry-Keating-Connes $xp$ model
and continued with several works aimed to provide  a physical realization of the Riemann zeros. The main steps  
in this  approach are: i) spectral realization of Connes's $xp$ model using the  Landau model
of an electron  in a magnetic field and electrostatic potential,  ii) construction of modified quantum $xp$ models whose 
spectra reproduce, in average, the behavior of the {\em zeros}, iii) reformulation of the $x(p + 1/p)$ model 
as  a  relativistic theory of a massive Dirac fermion in a region  of Rindler space-time, 
iv) inclusion  of the prime numbers into the massless Dirac equation by means of delta function potentials 
 acting  as moving mirrors that,  in the limit where they become semi transparent,  leads to  a  spectral realization of the {\em zeros},
 v)  a route for  proving  the Riemann Hypothesis,  and vi) proposal of an  interferometer that may provide
 an experimental observation of  the zeros of  the Riemann zeta function  and other Dirichlet $L$-functions. 

The P\'olya-Hilbert  (PH) conjecture was proposed as a physical explanation of  the RH
based on  the spectral  properties of self-adjoint operators:   there exists 
a {\em single}   quantum Hamiltonian containing  {\em all}  the Riemann zeros in its spectrum which are  therefore  real numbers. 
This  statement can  be called  the  {\em global} version of the PH conjecture. 
Instead of this, we have found  a  {\em local} version 
according to which a Riemann zero $E_n$  becomes  an eigenvalue of the Hamiltonian $H_\vartheta$ 
provided the  parameter $\vartheta$, that characterizes the self-adjoint extension, 
 is fine tuned to the combination   $\theta(E_n) + \frac{\pi}{2} {\rm sign } Z'(E_n)$. 
 In this sense the Hamiltonian  provides a physical realization of $\zeta(\frac{1}{2} + i t)$, 
and not only  of the Riemann-Siegel $Z$ function. We have  given arguments  for a proof of the RH
by contradiction: the existence of a {\em zero} off the critical line implies that
the eigenstates of $H_\vartheta$ are non normalizable in the discrete or continuum sense, which is impossible
since $H_\vartheta$ is a self-adjoint operator. 
These results  are obtained in the
limit where the mirrors become transparent  and assumes the convergence of some mathematical  
series that need to be analyzed more thoroughly.  Finally, we have proposed an interferometer
made of fermions propagating in a array of mirrors that may  yield  an  
experimental observation of the Riemann zeros in the Lab.

\vspace{1.5cm} 

\section*{Acknowledgements} I am   grateful for fruitful discussions and comments 
to Julio Andrade, Manuel Asorey,   Michael  Berry, Ignacio Cirac, 
Charles Creffield,  Jon  Keating,   Jos\'{e} Ignacio Latorre,  Giuseppe Mussardo,  Andr\' e LeClair,  Miguel Angel Mart\'{\i}n-Delgado, 
Javier Molina-Vilaplana,    
Javier  Rodr\'{\i}guez-Laguna, Mark Srednicki  and Paul  Townsend. I thanks  Denis Bernard for pointing 
out an error in the  first  version of this manuscript. 
I acknowledge
financial support from the grants FIS2012-33642,  FIS2015-69167-C2-1-P, 
QUITEMAD+ S2013/ICE-2801, and grants SEV-2012-0249, 
and SEV-2016-0597 of the "Centro de Excelencia Severo
Ochoa"  Program.




\begin{thebibliography}{999}
\bibitem{core} The expression for  ${\rm Re}_{s=0}  \; F(s)$  corresponding to the case $m=1$ contains the 
term $- \frac{1}{2}  \zeta''(z)/(\zeta'(z))^2$ which was in fact omitted  in the reference \cite{S14}. 
Nevertheless the same results are achieved.
\end{thebibliography}
\end{document}